\documentclass[preprint,12pt]{elsarticle}
\usepackage{graphicx}
\usepackage{bm}
\usepackage{chngcntr}
\usepackage{amssymb}
\usepackage{amsmath}
\usepackage{float}
\floatstyle{plaintop}
\restylefloat{table}
\usepackage[T1]{fontenc}

\usepackage{lineno}

\biboptions{authoryear}

\journal{Bull. Seismol. Soc. Am.}

\begin{document}

\begin{frontmatter}


\title{Spring-Slider and Finite Element Modeling of Microseismic Events and Fault Slip during Hydraulic Fracturing}



\author{Ali Kashefi$^{1}$, Eric M. Dunham$^{2}$, Benjamin Grossman-Ponemon$^{1}$, and Adrian J. Lew$^{1}$ }

\address{1. Department of Mechanical Engineering, Stanford University, Stanford, CA}
\address{2. Department of Geophysics, Stanford University, Stanford, CA}


\begin{abstract}


Hydraulic fracturing increases reservoir permeability by opening fractures and triggering slip on natural fractures and faults. While seismic slip of small faults or fault patches is detectable as microseismic events, the role of aseismic slip is poorly understood. From a modeling standpoint, geomechanical analysis using the Coulomb criterion can determine if faults slip but not whether slip is seismic or aseismic. Here we propose a computational methodology to predict fault slip, and whether slip is seismic or aseismic, using rate-and-state friction. To avoid computational costs associated with resolving small faults, we use the spring-slider idealization that treats faults as points. Interaction between faults is neglected. The method is applied to study fault slip from a hydraulic fracture that grows past a fault, without intersecting it. We represent the hydraulic fracture stressing using an asymptotic expansion of stresses around the tip of a tensile crack. We investigate the effect of fault length, orientation, and distance from the hydraulic fracture. For velocity-weakening faults with stiffness smaller than a critical stiffness, slip is seismic, whereas faults with stiffness greater than the critical stiffness slip aseismically. Furthermore, we compare the spring-slider idealization with a finite element analysis that resolves spatially variable slip. The spring-slider idealization provides reasonably accurate predictions of moment and even moment-rate history, especially for faults having stiffness close to or larger than the critical stiffness. Differences appear for large faults where rupture propagation is important, though differences might still be negligible for many applications. The spring-slider methodology could be applied to model statistics of microseismicity and aseismic slip on a population of small faults in a reservoir by populating a stochastic fracture network with spring sliders having frictional properties drawn from a statistical characterization based on well logs and experimental correlations between friction and rock properties.

\end{abstract}


\end{frontmatter}



\section*{Introduction}

Slip on small fractures generates microseismicity \citep{mckean2019quantifying}. Microseismic monitoring \citep[e.g., ][]{warpinski2009microseismic,maxwell2010petroleum,le2016hydraulic} and microseismic data analysis play a critical role both in geological sciences \citep[e.g., ][]{haring2008characterisation} and energy industries \citep[e.g., ][]{maxwell2011microseismic}. Microseismicity provides constraints on the fracture systems that are created or activated by hydraulic fracturing and related reservoir stimulation techniques used in the energy industry \citep{mcgillivray2004microseismic}, variation of stress fields in reservoirs \citep{le2016hydraulic}, and spatial and temporal mapping of hydraulic fracture growth \citep{baig2010microseismic}. Furthermore, microseismicity can reveal the transmission of pressure changes into and possible activation of hazard-scale faults \citep[see e.g.,][]{de2011geomechanical,holland2013earthquakes,atkinson2015ground}. From the perspective of understanding reservoir stimulation, one challenge is that microseismic events only reflect the seismic slip taking place within the reservoir. Several authors have argued for the important role that aseismic slip might play in the simulation process \citep{ mcclure2011investigation, zoback2012importance}. 


There are two general classes of models that have been used to study seismicity, and in particular microseismicity, in reservoirs and other systems involving hydraulic fractures (e.g., volcanic areas). The first class of models involves the explicit representation of small faults that host microseismic events, through the use of a boundary element method or a volume-discretized method like finite elements for solution of the mechanics problem in conjunction with some friction law. These methods provide predictions of seismicity on the explicitly resolved faults. The second class of models is more statistical in nature, and aims to predict seismicity rate changes throughout the system in response to imposed loading. Specific faults are not explicitly represented in this class of models.



Most studies fall within this first class of models that explicitly resolve faults \citep[e.g.,][]{mcclure2011investigation,dieterich2015modeling,prevost2016faults,roux2016microseismic,kroll2017sensitivity,li2019review,he2019study,sherman2019recovering}. \citet{dieterich2015modeling} and \citet{kroll2017sensitivity} utilized the RSQSim earthquake simulator (which combines boundary element solution of the elasticity problem with an approximation to rate-and-state friction) for the prediction of seismicity induced by fluid injection in reservoirs. \citet{prevost2016faults} simulated faults in reservoirs using an extended finite element method with the Coulomb failure criterion for slip. \citet{he2019study} modeled the response of a network of $\sim$100-m-long natural fractures to the stress changes from a growing hydraulic fracture in a 2D plane strain model, with slip-weakening friction on the natural fractures. Their results highlight the role of stress transfer between adjacent natural fractures in triggering slip far from the hydraulic fracture. \citet{sherman2019recovering} introduced a method to simulate microseismicity in coupled geomechanics and fluid flow simulations involving hydraulic fracture networks and faults in unconventional oil and gas reservoirs. Although this method could effectively predict the timing, location, and amplitude of the microseismic events, it could not discriminate between aseismic and seismic slip due to use of the Coulomb failure criterion. Rate-and-state friction was coupled to geomechanics and fluid flow along fractures in the pioneering work of \citet{mcclure2011investigation}. Quasi-static stress interactions were determined by boundary element solution of the elasticity problem, as in \citet{dieterich2015modeling} and \citet{kroll2017sensitivity}.




The second class of models predicts seismicity rate changes, primarily through the use of a theory for earthquake nucleation rates based on rate-and-state friction. This theory was developed by \citet{dieterich1994constitutive} and elaborated upon by \citet{heimisson2018constitutive} and \citet{heimisson2019constitutive}. It has been used to connect with seismicity data recorded during dike intrusions (growth of magma-filled hydraulic fractures). Stress changes from dike opening were used to predict seismicity rates, which were then combined with crustal deformation measurements to infer the time-dependent geometry and opening of the dike \citep{segall2013time,heimisson2019fully}. The primary disadvantage with this approach is that the \citet{dieterich1994constitutive} theory only predicts the rate of nucleation, not the eventual size of the earthquake. 


By and large, there have been two fundamental challenges in this area. First, most of these methods (with a few exceptions) utilize the Coulomb failure criterion and hence cannot distinguish between seismic and aseismic slip. Second, approaches involving explicit discretization of small faults have extremely high computational costs because high resolution meshes are needed to resolve deformation around these faults.

To obviate the aforementioned problems and overcome these two challenges, we propose a novel method for microseismicity simulation. The method combines rate-and-state friction, a widely used frictional framework that captures the diversity of sliding behaviors seen experimentally, with the spring-slider idealization of the response of an elastic solid containing a small fault. Hereafter we utilize the term fault to refer specially to the fault patch or natural fracture experiencing slip as a consequence of stress loading, with the term fracture reserved for the hydraulic fracture that provides that stress loading. 

Rate-and-state friction \citep{dieterich1979modeling,ruina1983slip, RiceRunio1983, marone1998laboratory, rice2001rate} captures the dependence of friction coefficient on slip velocity, in particular whether friction increases or decreases with an increase in slip velocity (i.e., velocity-strengthening or velocity-weakening behavior). Moreover, the velocity dependence of friction, and other rate-and-state parameters, can be experimentally constrained. In the context of the hydraulic fracture application, we point out the work of \citet{kohli2013frictional} that provides rate-and-state friction parameters for shale reservoir rocks, with systematic trends in velocity dependence and friction coefficient as a function of clay plus organic content.

The spring-slider idealization \citep[e.g.,][]{dieterich1992fault, scholz1998earthquakes} captures the elastic response of the rock to fault slip in a computationally efficient manner by suppressing all spatial dependence of slip, slip velocity, and stresses on the fault. The coupled frictional spring-slider equations depend only on time and can be solved with negligible expense (as compared to finite elements or other spatial discretization techniques) with adaptive time integration. The overall method quantifies slip, slip velocity, stress drop, and earthquake magnitude, and hence is a valuable tool for microseismic monitoring.

Our objectives in this study are twofold: first, to demonstrate how the spring-slider model can be used to investigate controls on microseismicity and aseismic slip during hydraulic fracturing; and second, to assess the accuracy and reliability of the spring-slider predictions by comparison to a finite fault model. These challenges are investigated in the context of the idealized problem of a small fault on which slip is triggered by the stress changes caused by the passage of a hydraulic fracture. The fracture propagates in a straight line without intersecting the fault, and pore pressure changes from leak-off and poroelasticity are neglected. The spring-slider model is not restricted to this idealized problem, of course, and could be incorporated into more sophisticated and realistic 3D geomechanical models that capture all of the complexities involved in hydraulic fracturing \citep[see e.g.,][]{lele2019modeling}. For the first objective, we analyze the slip and slip velocity (i.e., moment and moment rate) and stress drop predicted by the spring-slider model for faults of various lengths, orientations, and distances from the hydraulic fracture. We focus specifically on the distinction between seismic and aseismic slip, illustrating how the classical concept of a critical fault stiffness predicts the slip behavior. For the second objective, we compare the spring-slider predictions to those from a finite element simulation that resolves the finite-length fault through solution of the 2-D elasticity equations coupled to rate-and-state friction on the fault. 


\section*{Problem Formulation}\label{ProblemFormulation}
To begin, we state the 2-D plane strain problem that is solved in this study. A hydraulic fracture approaches a fault in a homogeneous, linear elastic solid whole space without intersecting the fault (Fig. \ref{Fig1}). The solid has maximum and minimum principal stresses $\sigma_1$ and $\sigma_3$, and the hydraulic fracture advances in a straight line in the direction of $\sigma_1$ at a constant velocity $U$. The pressure of fluid inside the fracture is equal to $p$, consistent with the fracture propagating in the toughness-dominated limit \citep{garagash2000tip} with near-tip suctions and lag confined to a region that is smaller than other length scales of interest (e.g., distance between the hydraulic fracture and fault). Adjacent to the hydraulic fracture is a planar fault of length $L$ and orientation $\alpha$ with respect to the path of the hydraulic fracture. The normal distance between the hydraulic fracture path and the fault center is $H$. Pore pressure changes and poroelastic effects are neglected.

Rather than solving the fully coupled problem that accounts for two-way interactions between the hydraulic fracture and fault, we assume one-way coupling from the hydraulic fracture to the fault. Specifically, the hydraulic fracture creates stress changes in the solid that load the fault, potentially triggering fault slip. However, the stress changes caused by fault slip are only taken into account when solving the faulting problem, and have no influence on the growth or opening of the hydraulic fracture and the stresses around its tip. This is a valid approximation when stress changes from fault slip are much smaller than other stress changes driving hydraulic fracture growth (e.g., $p-\sigma_3$). This is likely the case due to the very small size of microseismic events, as discussed further in the {\bf Discussion and Conclusions} section, where we also discuss conditions for neglecting interactions between multiple microseismic events.

We compare two models for fault slip: 1.) a spring-slider model in which stress loading from the hydraulic fracture is evaluated at the fault center and the finite length of the fault is captured only through spring stiffness, and 2.) a finite fault model, solved with the finite element method, that resolves spatially variable loading and slip dynamics along the fault. 

\subsection*{Stress Field Around The Tip of a Hydraulic Fracture}\label{StressFieldAroundTheTip}
In this study, we use a simple analytical expression for the stresses around the hydraulic fracture. Anticipating that fault slip will be triggered by the stress concentration near the hydraulic fracture tip, we utilize the asymptotic expansion for a singular mode I crack. Assuming 2-D plane strain deformation, the near-tip stress is written as \citep[e.g., ][]{freund1998dynamic}
\begin{equation}
\begin{bmatrix}
\sigma_{xx} & \sigma_{xy} \\
\sigma_{xy} & \sigma_{yy} 
\end{bmatrix}
=
\begin{bmatrix}
-\sigma_{1} & 0 \\
0 & -p
\end{bmatrix}
+\frac{K_I}{\sqrt{2\pi r}}
\begin{bmatrix}
f_{11}(\theta) & f_{12}(\theta) \\
f_{12}(\theta) & f_{22}(\theta)
\end{bmatrix},
\label{Eq1}
\end{equation}
with

\begin{equation}
f_{11}(\theta)=\cos{\frac{\theta}{2}}\left(1-\sin{\frac{\theta}{2}}\sin{\frac{3\theta}{2}}\right),
\label{Eq2}
\end{equation}

\begin{equation}
f_{12}(\theta)=\cos{\frac{\theta}{2}}\sin{\frac{\theta}{2}}\cos{\frac{3\theta}{2}},
\label{Eq3}
\end{equation}

\begin{equation}
f_{22}(\theta)=\cos{\frac{\theta}{2}}\left(1+\sin{\frac{\theta}{2}}\sin{\frac{3\theta}{2}}\right),
\label{Eq4}
\end{equation}
where $K_I$ is the stress intensity factor, and $(r,\theta)$ are polar coordinates centered at the moving hydraulic fracture tip. The distance from the hydraulic fracture tip to the center point of the fault (where the spring-slider is positioned) is 
\begin{equation}
r=\sqrt{(Ut)^2+H^2}
\label{Eq5},
\end{equation}
where $t$ is time, defined so that the hydraulic fracture passes the fault center at $t=0$. The finite fault model accounts for spatially variable loading across the fault. We denote some position on the fault as $\textbf{\textit{x}}_{fault}$ with respect to the origin of coordinate system, located at the hydraulic fracture tip position at $t=0$. For the spring-slider model, $\textbf{\textit{x}}_{fault}$ refers to the fault center.

We introduce the nondimensional time
\begin{equation}
t^{*}=\frac{t}{H/U},
\label{Eq6}
\end{equation}
which also serves to quantify the $x$-position of the hydraulic fracture and the fault ($Ut$) as compared to $H$. We use $t^{*}$ in our plots in the future sections. Furthermore, we nondimensionalize stress as $\sigma_{ij}^*={\sigma_{ij}}/{p}$ and the stress intensity factor as $K_I^*={K_I}/{p\sqrt{2 \pi H}}$.

\subsection*{Coulomb Failure Criterion and a Graphical Example}\label{CoulombFailureCriterion}

To illustrate how hydraulic fracture stressing can trigger slip, we introduce the Coulomb failure criterion (which we later replace with rate-and-state friction). In the absence of cohesion, slip occurs when the resolved shear and normal tractions on the fault, $\tau(\textbf{\textit{x}}_{fault},t)$) and $\sigma(\textbf{\textit{x}}_{fault},t)$, respectively, first satisfy
\begin{equation}
\tau(\textbf{\textit{x}}_{fault},t) = f_0 \sigma(\textbf{\textit{x}}_{fault},t),
\label{Eq11}
\end{equation}
where $f_0$ is the reference friction coefficient and $\sigma>0$ in compression.

We select parameter values $f_0=0.6$ \citep[a standard choice for friction in crystalline rocks,][]{dieterich1981constitutive}, $\sigma_1^*=2.0$, and $K_I^*=0.5$. The stress state $\sigma_1^*=2.0$ is close to but slightly below failure (which occurs at $\sigma_1^*=3.1$ when the Coulomb failure lines become tangent to Mohr’s circle). The dimensionless stress intensity factor $K_I^*=0.5$ is sufficiently large to trigger slip for some but not all fault orientations, thus providing a nontrivial example of forcing. Given these parameter choices, Figure \ref{Fig2}a depicts the stress evolution at the center of a fault oriented at $\alpha=35^{\circ}$. The Coulomb failure criterion is plotted as a dashed black line. Figure \ref{Fig2}c shows the hydraulic fracture in four different states. In state (1), the hydraulic fracture is far to the left of the fault and stress lies exactly on Mohr’s circle (Fig. \ref{Fig2}a, state (1)). The hydraulic fracture then approaches the fault (Fig. \ref{Fig2}a and c, state (2)), causing a normal stress reduction with relatively no change in shear stress. As the hydraulic fracture comes closer to the fault, and passes directly beneath its center in state (3), the stress state at the center of the fault moves into the Coulomb failure region. After the hydraulic fracture has passed the fault (state (4) in Fig. \ref{Fig2}a and c), the stress moves out of the failure region.

Figure \ref{Fig2}a and Figure \ref{Fig2}c explain the stress trajectory for a fault with $\alpha=35^{\circ}$; however, one may repeat the same exercise for other orientations as we do in Fig. \ref{Fig2}b. For each trajectory, our concern is if the fault slips (which can be answered by the Coulomb failure criterion) and, if so, whether it slips seismically or aseismically (which cannot be answered by the Coulomb failure criterion). To assess the stability of slip, we must account for elastic stress changes in response to slip and a more realistic friction law.

\subsection*{Elastic Response to Slip and the Spring Slider Idealization}\label{SpringSliderIdealization}

Here we describe the elastic stress changes from fault slip and how those can be approximated with the spring-slider model. Slip $\delta$ is defined as the discontinuity in tangential displacement across the fault. We further enforce no opening across the fault and require tractions on opposing sides of the fault to be equal and opposite. We then exploit linearity of the elastic solid surrounding the hydraulic fracture and fault to write the stresses $\sigma_{ij}$ as the superposition of loading stresses from the hydraulic fracture, $\sigma_{ij}^{HF}$, and stress changes from fault slip, $\Delta \sigma_{ij}$:
\begin{equation}
    \sigma_{ij} = \sigma_{ij}^{HF} + \Delta \sigma_{ij}
\end{equation}
The slip-induced stress changes can be further decomposed into quasi-static stress changes, $\Delta \sigma_{ij}^{QS}$, obtained from solving the static elasticity problem with imposed slip, and dynamic stress changes associated with elastic waves. As in many other earthquake modeling studies \citep{erickson2020community}, we approximate the dynamic stress changes using the radiation-damping approximation \citep{rice1993spatio} as described subsequently. The shear and normal tractions on the fault, $\tau(\textbf{\textit{x}}_{fault},t)$ and $\sigma(\textbf{\textit{x}}_{fault},t)$, are obtained from $\sigma_{ij}$, and can also be written as a superposition of loading and stress change from slip.

The quasi-static stress change, $\Delta \sigma_{ij}^{QS}$, is a nonlocal, linear mapping of slip and, for spatially variable slip, can be computed using a variety of numerical methods. In this work, we use finite elements, which requires truncating the computational domain at a finite distance from the fault (details provided in the Appendix).

Sufficiently small faults can be idealized as point sources through the spring-slider model \citep[e.g., ][]{dieterich1992earthquake}. Assuming a spatially uniform change in shear stress on the fault, the quasi-static elastic response to slip is given by the classic mode II crack solution of \citet{starr1928slip}. This solution provides an elliptical slip distribution, and in the spring-slider idealization the linear relation between quasi-static shear stress change and slip is obtained by evaluating this solution in the center of the fault. The resulting stress change is written as $-k \delta$, where $\delta$ is slip in the fault center and the stiffness $k$ is
\begin{equation}
k = \frac{G}{(1-\nu)L},
\label{Eq22}
\end{equation}
where $G$ is the shear modulus and $\nu$ is the Poisson ratio. Inertial effects associated with wave radiation are captured using the radiation-damping approximation \citep{rice1993spatio}, which adds to the shear stress change the term $-\eta V$, where 
\begin{equation}
V = \frac{d\delta}{dt}
\label{Eq21}
\end{equation}
is the slip velocity and
\begin{equation}
\eta = \frac{G}{2c_s}
\label{Eq23}
\end{equation}
is the impedance of a pair of plane shear waves, propagating at speed $c_s$, radiating from the fault.

Consequently, the shear stress on the fault in the spring-slider model can be written as
\begin{equation}
\tau = \tau^{HF} - k\delta - \eta V,
\label{Eq20}
\end{equation}
where $\tau^{HF}$ is evaluated at $\textbf{\textit{x}}_{fault}$, the position of the fault center. A similar expression is used for finite fault modeling, except that $-k \delta$ is replaced with the quasi-static elastic stress change, $\Delta \tau^{QS}$, computed with finite elements, and the expression is enforced at each point on the fault.

The normal stress change from slip on a planar fault in a uniform whole space is zero \citep{starr1928slip}, such that the normal stress is simply the resolved normal loading from the hydraulic fracture:
\begin{equation}
\sigma(\textbf{\textit{x}}_{fault},t)= \sigma^{HF}(\textbf{\textit{x}}_{fault},t).
\label{Eq28}
\end{equation}
This expression is utilized in spring-slider modeling; however, due to truncation of the computational domain in our finite fault model, the normal stress change from slip is nonzero (but small) and we account for it when coupling to friction.

\subsection*{Rate-and-State Friction}\label{RateAndStateFrictionLaw}
With rate-and-state friction \citep[e.g., ][]{ruina1983slip,RiceRunio1983,marone1998laboratory,rice2001rate}, the shear stress on the fault ($\tau$) is always equal to the frictional shear strength ($f \sigma$):
\begin{equation}
\tau= f \sigma.
\label{Eq25}
\end{equation}
The friction coefficient $f$ is a function of slip velocity ($V$) and a state variable ($\psi$) and can be written as \citep[e.g.,][]{rice2001rate}:
\begin{equation}
f(V,\psi)=a\sinh^{-1}\left(\frac{V}{2V_0}\exp\left(\frac{\psi}{a}\right)\right),
\label{Eq26}
\end{equation}
where $a$ is the direct effect parameter. The reference slip velocity is $V_0$ at which the steady state friction coefficient is $f_0$. The state variable $\psi$ captures the past history of sliding and quantifies changes in interface contact strength. We use the slip law for state evolution \citep[e.g.,][]{rice2001rate}:

\begin{equation}
\frac{d\psi}{dt}=-\frac{V}{d_c}(f-f_{ss}),
\label{Eq27}
\end{equation}
with steady state friction coefficient
\begin{equation}
f_{ss}=f_0-(b-a)\ln\left(\frac{V}{V_0}\right),
\label{Eq27_1}
\end{equation}
where $d_c$ is the state evolution distance and $a-b$ quantifies the velocity dependence of friction in steady sliding. The fault is velocity strengthening for $a-b>0$ and velocity weakening if $a-b<0$. In this study, we restrict attention to velocity-weakening friction, for which sliding can become unstable, and this is the case considered in the numerical simulations later; however, the method we develop can equally well be applied with velocity-strengthening friction. 

For a spring-slider system with velocity-weakening friction, steady sliding is unstable to perturbations when the stiffness $k$ is smaller than the critical stiffness \citep{RiceRunio1983}
\begin{equation}
k_{cr} = \frac{\sigma(b-a)}{d_c}.
\label{Eq24}
\end{equation}
Thus if $k<k_{cr}$, slip occurs seismically, whereas if $k>k_{cr}$, slip is aseismic. Because $k$ is inversely proportional to fault length $L$ (equation (\ref{Eq22})), one can interpret the instability condition $k<k_{cr}$ as $L > L_{cr}$ where $L_{cr}$ is a critical length.

\subsection*{Coupling Rate-and-State Friction and Elasticity}

In this section we discuss how rate-and-state friction is coupled to elasticity, which leads to a dynamical system for slip $\delta$ and state variable $\psi$. Elasticity enters the problem when determining slip velocity $V$, which is used to update $\delta$ and $\psi$. We begin with the spring-slider model, inserting the elastic stress changes given by equations (\ref{Eq20}) and (\ref{Eq28}) into the rate-and-state friction law, equation (\ref{Eq25}), to yield
\begin{equation}
\tau^{HF}-k\delta-\eta V= f(V,\psi)\sigma^{HF}.
\label{Eq29}
\end{equation}
This equation can be solved for slip velocity $V$ given values of slip $\delta$ and state variable $\psi$.

A similar approach is taken for finite fault modeling with finite elements. Again, we exploit linearity of the elasticity equations to write stress as the superposition of loading from the hydraulic fracture and stress changes caused by slip. The finite element code provides the quasi-static shear and normal stress changes (via a linear, but nonlocal, mapping of fault slip), $\Delta \tau^{QS}$ and $\Delta \sigma^{QS}$, respectively, and we add the radiation-damping term to the quasi-static shear stress change. Inserting these expressions into the rate-and-state friction law, equation (\ref{Eq25}), yields
\begin{equation}
\tau^{HF} + \Delta \tau^{QS} - \eta V = f(V,\psi)(\sigma^{HF} + \Delta \sigma^{QS}).
\label{Eq32}
\end{equation}
This equation is solved for $V$ at each node on the fault. Coupling between different points on the fault is provided by the nonlocal relation between fault slip $\delta$ and the quasi-static stress changes $\Delta \tau^{QS}$ and $\Delta \sigma^{QS}$.

The governing equations (after spatial discretization for the finite fault case) can be viewed as nonlinear ordinary differential equations in time for slip and state, subject to a nonlinear algebraic constraint (equation (\ref{Eq29}) for the spring-slider and equation (\ref{Eq32}) for the finite fault model). Slip and state are integrated in time using an adaptive explicit Runge-Kutta scheme, specifically the 3(2) Runge-Kutta pair \citep{hairer2011solving}, with error control on both slip and state. At each time step, we solve equation (\ref{Eq29}) or equation (\ref{Eq32}) for the slip velocity $V$ using a bracketed Newton-Raphson method \citep{kozdon2012interaction}. The state rate is obtained from equation (\ref{Eq27}). Then we integrate equations (\ref{Eq27}) and (\ref{Eq21}) for $\psi$ and $\delta$ using the Runge-Kutta method. One may refer to \citet{hetland2010post}, \citet{erickson2014efficient}, \citet{allison2018earthquake}, and \citet{torberntsson2018finite} for further details of the time integration algorithm.
 
\section*{Results}\label{ResultsAndDiscussion}
\subsection*{Spring-Slider Modeling of Microseismic Events and Fault Slip during Hydraulic Fracturing}\label{Spring-sliderModelingOfMicroseismicEvents}

In this section we apply the spring-slider model to study microseismic events and fault slip during hydraulic fracturing. Specifically, we investigate the effect of fault length $L$, fault orientation $\alpha$, and the distance $H$ between the hydraulic fracture path and the fault center. The material parameters are $G=30$ GPa, $\nu=0.25$, and $\eta=4.41$ MPa/(m/s). The frictional parameters are $a=0.015$, $b=0.20$, $V_0=10^{-6}$ m/s, $f_0=0.6$, and $d_c=10^{-5}$ m. The initial state variable is $\psi=0.606$ and the initial slip on the fault is $\delta=0$. All values are chosen for consistency with \citet{torberntsson2018finite}. Simulations are started from approximately $t^*=-13.5\times 10^4$, though the exact starting time has negligible influence on the results provided that the initial $t^*$ is sufficiently small. The fluid pressure is set to $p=30$ MPa, a reasonable choice for hydraulic fracturing at one or a few kilometers depth. The maximum principle stress is $\sigma_1=75$ MPa and the stress intensity factor is $K_I=25$ MPa$\sqrt{\textrm{m}}$. Note that $\sigma_3$ does not appear explicitly in our problem formulation; however, $\sigma_3$ affects the choice of $K_I$ because for height-bounded hydraulic fractures the stress intensity factor scales as $(p-\sigma_3)\sqrt{h}$, where $h$ is the height of the fracture \citep[see e.g.,][]{green1950distribution, perkins1961widths,nordgren1972propagation,zoback2019unconventional}. Typical net pressures $p-\sigma_3$ of a few MPa and fracture height of order 10 m \citep[e.g.,][]{valko1994propagation} gives a value of $K_I$ like the one chosen. We later perform a few additional simulations to explore other values of $\sigma_1$ and $K_I$. Finally note that the approach of the hydraulic fracture toward the fault causes a very small amount of aseismic slip. Thus, in the plots to follow, and our calculations of seismic moment and magnitude, this aseismic slip is neglected. Specifically, the reported slip is defined with respect to the aseismic slip at $t^*=-2$.

While the spring-slider results are presented in terms of slip and slip velocity, these are best interpreted as moment and moment rate. Seismic moment is related to slip $\delta$ (which in the spring-slider model is the maximum slip at the fault center) and fault area $S$ as
\begin{equation}
M_0=\frac{2}{3}G\delta S.
\label{Eq33}
\end{equation}
While the specific relation between spring-slider stiffness and fault length $L$ comes from a 2-D plane strain crack solution (to facilitate comparison with the 2-D finite fault model), in reporting moment (and magnitude) we approximate $S$ as a circle with the diameter of $L$. The seismic moment is therefore approximated as 
\begin{equation}
M_0\approx\frac{\pi}{6}G\delta L^2.
\label{Eq34}
\end{equation}
Given $M_0$, the magnitude is calculated as \citep{hanks1979moment}
\begin{equation}
M_{w}=\frac{2}{3}(\log_{10}(M_0)-9).
\label{Eq35}
\end{equation}

First, we set $\alpha=35^{\circ}$ and $H=1.3$ m and explore two different fault lengths: $L=1.3$ m and $L=0.4$ m. For $L=1.3$ m, $k<k_{cr}$ and seismic slip is expected, whereas for $L=0.4$ m, $k>k_{cr}$ and aseismic slip is expected. This is exactly what we observe (Fig. \ref{Fig4}). For both cases, the fault slips when the hydraulic fracture tip is very close to, but still approaching, the fault center. Using Fig. \ref{Fig4}a-b the predicted magnitudes, $M_w=-1.39$ and $M_w=-2.94$ for the seismic and aseismic cases, respectively, are in an acceptable range of $-3$ to $-1$ for microseismic event data \citep[see e.g.,][]{shemeta2010s, warpinski2012measurements}. Similarly, the stress drop (Fig. \ref{Fig4}e and f) is roughly equal to 6.0 MPa and 0.9 MPa for seismic and aseismic events, respectively, in good agreement with typical stress drops (0.1 to 10 MPa) reported in the literature \citep{kwiatek2011source} for microseismic events. The agreement in stress drop and magnitude comes from having selected appropriate frictional and stress parameters as well as reasonable fault lengths.

Second, we explore the role of fault orientation. Fault orientation is well known to play an important role in seismicity studies because it determines the closeness of the fault to frictional failure \citep[e.g.,][]{zoback2012importance}. To study orientation, we fix $L=1.3$ m and $H=1.3$ m and vary $\alpha$ from 0 to 90$^{\circ}$. Results are shown in Fig. \ref{Fig5}. The fault experiences no slip for $0^{\circ} \leq \alpha \leq 16^{\circ}$. In a narrow range of $17^{\circ} \leq \alpha \leq 18^{\circ}$, the fault slips aseismically. Seismic slip occurs for $19^{\circ} \leq \alpha \leq 36^{\circ}$. For $37^{\circ} \leq \alpha \leq 46^{\circ}$, the fault slips aseismically. Eventually, the fault does not slip in the range of $47^{\circ} \leq \alpha \leq 90^{\circ}$. Note that while there is a continuous change in maximum slip velocity with $\alpha$, we utilize the terms seismic slip, aseismic slip, and no slip based on the following thresholds: seismic slip has maximum $V$ greater than $10^{-3}$ m/s, aseismic slip has maximum $V$ between $10^{-9}$ m/s and $10^{-3}$ m/s, and no slip has maximum $V$ below $10^{-9}$ m/s.

We examine this phenomenon more closely by focusing on $\alpha=36^{\circ}$. The associated stress trajectory, plotted in pink in Fig. \ref{Fig5}a, violates the Coulomb failure criterion. The slip velocity history (Fig. \ref{Fig5}b, pink), shows that the fault slips seismically. Next we increase the fault orientation by one degree to $\alpha=37^{\circ}$. Figure \ref{Fig5}a shows a similar stress trajectory for $\alpha=37^{\circ}$ (cyan) that, like the $\alpha=36^{\circ}$ trajectory, violates Coulomb failure. However, the fault slips aseismically (Fig. \ref{Fig5}b, cyan).

Third, we examine the distance $H$ between the fault and hydraulic fracture. This is done by fixing $L=1.3$ m and $\alpha=35^{\circ}$ and then increasing $H$ from 1.3 m to 6.0 m. To avoid intersection of the fault and hydraulic fracture, the minimum $H$ is $L\sin(\alpha)/2$. Figure \ref{Fig6} shows the evolution of slip velocity and slip for different values of $H$. Fig. \ref{Fig6}a shows a transition from seismic slip at small $H$ to aseismic slip at $H=1.5$ m. The amount of aseismic slip decreases continuously as $H$ increases.

Building on these three specific case studies, we next perform a more comprehensive investigation of $\alpha \in [0^{\circ}, 180^{\circ}]$ and $H \in [1.3 \textrm{ m}, 2.6 \textrm{ m}]$. Looping through each parameter combination, we run a spring-slider simulation and take the maximum absolute value of slip and slip velocity and compute the magnitude. Figure \ref{Fig7} provides these results. First note that significant slip is limited to a range of angles centered around the conjugate critical orientations from a standard Mohr-Coloumb analysis in the spatially uniform background stress state. However, the hydraulic fracture loading breaks the symmetry between the conjugate orientations, favoring seismic slip for $\alpha$ around 30$^{\circ}$. Triggered slip also occurs for orientations around 150$^{\circ}$, but this slip tends to be aseismic though it persists for larger $H$. 


Thus far our simulations have been performed for fixed values of $\sigma_1$, $p$, and $K_I$. To show how these parameters might affect the results, we execute a few additional simulations in which these parameters are varied. Figure \ref{Fig7_1} shows maximum slip velocity as a function of $H$ and $\alpha$ for three different values of $\sigma_1/p$: 2.3, 2.5, and 2.9 with $p$ and $K_I$ fixed to their reference values. As expected, the higher levels of deviatoric stress in the solid (at higher $\sigma_1/p$) facilitate fault slip. Similarly, slip is favored at greater $H$ as $K_I$ is increased (Figure \ref{Fig7_2}). Notice that although by increasing $K_I$ or $\sigma_1/p$ the magnitude of maximum slip velocity increases over the $H-\alpha$ map in general, different angles experience different patterns for the maximum slip velocity. For instance, note the different response for orientations around $\alpha=30^{\circ}$ and 150$^{\circ}$: while slip becomes increasingly more seismic for $\alpha=30^{\circ}$ as $K_I$ or $\sigma_1/p$ is increased, it becomes less seismic for $\alpha=150^{\circ}$. The reason is that for $\alpha=150^{\circ}$, normal stress changes (as a result of an increase in $K_I$ or $\sigma_1/p$) decrease $k_{cr}$ (and thus the difference between $k$ and $k_{cr}$); however, this is not the case for $\alpha=30^{\circ}$.


Overall, we find that hydraulic fracture stressing only triggers fault slip within a few meters of the hydraulic fracture tip. The triggering distance increases as background deviatoric stress increases, but never approaches the $\sim$100 m triggering distances that are routinely reported in microseismic monitoring studies. This suggests that triggering at greater distances must come from some other processes neglected in our model, such as pore pressure changes from leak-off into the natural fracture system of the reservoir.

\subsection*{Comparison Between Spring-Slider and Finite Fault Modeling of Microseismic Events during Hydraulic Fracturing}\label{AComparisonBetweenSpring-sliderAndFiniteElementModeling}

In this section, we compare the spring-slider model with a 2D finite fault model computing with finite elements. For this comparison, we fix the fault length $L=1.0$ m (consequently fixing the stiffness $k$) and we vary the state evolution distance $d_c$ and consequently the critical stiffness $k_{cr}$. We select $d_c$ to examine three different cases: $k>k_{cr}$, $k\approx k_{cr}$, and $k<k_{cr}$. We set $H=1.3$ m and $\alpha=35^{\circ}$, and all other parameters are the same as in the previous section.

The computational domain is the square $\Omega := [0 \textrm{ m}, 10 \textrm{ m}]\times[0 \textrm{ m}, 10 \textrm{ m}]$. The fault is at the center of the domain, parallel to the $x$-axis. Zero displacement boundary conditions are enforced on all sides of the square. The mesh has 22910 nodes and 45416 elements (Fig. \ref{Fig3}b), with small elements clustered around the fault and larger elements away from it. The large domain size as compared to the fault length mimics the whole-space assumption used in deriving the spring-slider stiffness, and makes results fairly insensitive to the boundary conditions on the exterior of the domain.

Figures \ref{Fig8}, \ref{Fig10}, and \ref{Fig12} illustrate the evolution of slip ($\delta$), slip velocity ($V$), and shear stress ($\tau$) on the fault for the state evolution distance of $d_c=14\times10^{-5}$ m, $2.4\times10^{-5}$ m, and $1.0\times10^{-5}$ m. Since an adaptive time stepping algorithm is used, we show the evolution of slip, slip velocity, and shear stress along the fault based on the time-step index rather than real time in these figures. Because earthquakes occur extremely fast, the time history of these quantities can be seen more easily with this type of plot.

We start with $d_c=14\times10^{-5}$ m, for which $k>k_{cr}$ and slip is expected to be aseismic. Figure \ref{Fig8} shows an approximately elliptical distribution of slip and uniform distribution of shear stress, consistent with the fault being smaller than the nucleation length. Slip is indeed aseismic with $V$ only reaching a few $\mu$m/s. Slip and slip velocity at the fault center, and the spatial average of shear stress, are plotted in Fig. \ref{Fig9}, along with predictions of the spring-slider model. There is excellent agreement between the spring-slider and finite fault models. The predicted magnitudes, $M_w=-2.24$ and $M_w=-2.21$ for the spring-slider and finite fault models, respectively, have only about 1\% error.

Next we decrease the state evolution distance to $d_c=2.4\times10^{-5}$ m, for which $k\approx k_{cr}$. Results are shown in Fig. \ref{Fig10}. In comparison with the previous case, the shear stress along the fault is more spatially variable and there is a sense of nucleation on the left side of the fault (closest to the hydraulic fracture) with gentle propagation to the right. The maximum slip velocity has increased by two orders of magnitude to $10^{-4}$ m/s, indicating the start of a transition from aseismic to seismic slip. There is again excellent agreement with the spring-slider model (Fig. \ref{Fig11}). 

Finally we decrease the state evolution distance to $d_c=1.0\times10^{-5}$ m, for which $k<k_{cr}$ and we expect seismic slip. Fig. \ref{Fig12} shows a clear distinction between nucleation and propagation, with the maximum slip velocity of 0.7 m/s at the rupture front. There are now more significant differences between the spring-slider and finite fault solutions (Fig. \ref{Fig13}). The earthquake nucleates later in the spring-slider model than in the finite fault model ($t^*=-0.23$ versus $t^*=-0.60$), likely because hydraulic fracture stressing is evaluated in the fault center for the spring-slider model. Despite these differences, the spring-slider model still accurately predicts the seismic nature of slip and the magnitude with relatively small error.


In summary, we find that while some quantitative differences are present, in general there is good agreement between the two approaches in terms of predicting the overall amount of slip (i.e., earthquake magnitude) and whether slip is seismic or aseismic. The differences between the two approaches increase as stiffness becomes much smaller than the critical stiffness, so that nucleation is confined to a small section of the overall fault, and slip takes the form of a propagating rupture.


From a computational cost point of view, decreasing $d_c$ increases the number of time steps and hence the computational cost. For instance, for $d_c=14\times10^{-5}$ m the finite element simulation takes 296 time steps, whereas for $d_c=2.4\times10^{-5}$ m it takes 599 time steps, and finally for $d_c=1.0\times10^{-5}$ m it takes 2903 time steps. Lastly, we emphasize the extreme efficiency of the spring-slider model relative to the finite element simulation, with the former taking just a few seconds and the latter a few hours in our present implementation.

\subsection*{Comparison Between Rate-And-State Friction and Coulomb Failure Criterion}\label{ComparisonBetweenRSFandCC}

We have focused thus far on rate-and-state friction, which is widely used in the earthquake modeling community. However, most geomechanical analyses utilize the simpler Coulomb failure criterion. Here we provide a comparison between these two approaches in the context of spring-slider modeling with radiation damping. Coupling the Coulomb failure criterion with the spring-slider model is straightforward; the sole modification to the numerical algorithm in section {\bf Coupling Rate-and-State Friction and Elasticity} comes when computing slip velocity. Instead of solving equation (\ref{Eq29}), we determine $V$ as
\begin{equation}
V=
\begin{cases}
\frac{\tau^{HF}-k\delta-f_0\sigma^{HF}}{\eta}  \textrm{  if  } \tau^{HF}-k\delta>f_0\sigma^{HF},\\
0,  \textrm{  otherwise.  }
\end{cases}
\label{Eq35_1}
\end{equation}

Figure \ref{Fig13_1} compare the evolution of slip, slip velocity, and shear stress computed by the rate-and-state friction law and the Coulomb failure criterion for faults with $L=0.4$ m ($k>k_{cr}$, aseismic) and 1.3 m ($k<k_{cr}$, seismic). The rate-and-state reference friction coefficient $f_0 = 0.6$ is also used for the Coulomb criterion.

As can be seen from Fig. \ref{Fig13_1}a, the amount of slip predicted with the Coulomb criterion is three times greater than that predicted with rate-and-state friction. This is because for $k>k_{cr}$, the rate-and-state frictional response is dominated by the direct effect, which is velocity-strengthening in character. This raises the friction coefficient above $f_0$ during sliding, so the rate-and-state model has more resistance and hence less slip than the Coulomb model. Nonetheless, the peak slip velocity is comparable for the two models.

In contrast with the aseismic event, the Coulomb failure criterion makes quite different predictions for the seismic event. Slip in the Coulomb model is entirely aseismic, due to the fact that sliding occurs at a constant friction coefficient. In contrast, because $k<k_{cr}$, the fault in the rate-and-state model experiences weakening toward the lower steady state friction coefficient. This frictional weakening process is unstable and takes the form of sliding that rapidly accelerates to the inertially limited peak slip velocity of order 1 m/s (Fig. \ref{Fig13_1}). For this reason, the Coulomb model underpredicts stress drop and final slip, and overpredicts the duration of slip.

Another important difference between these two models is the prediction of earthquake timing. As can be seen from Fig \ref{Fig13_1}, the Coulomb criterion predicts that sliding begins at approximately $t^*=-1.24$ for both the aseismic and seismic events, whereas the onset of significant sliding with rate-and-state friction is approximately $t^*=-0.6$ and $t^*=-0.3$ for the aseismic and seismic events, respectively.

Overall, while the Coulomb failure criterion is effective at determining whether or not fault slip occurs, it is ineffective at accurately predicting the timing of slip and the slip history, especially for seismic slip cases. For seismic slip it is essential to account for frictional weakening and stress drop, which cause the earthquake instability.

\section*{Discussion and Conclusions}\label{Conclusions}

In this work, we have applied the spring-slider model with rate-and-state friction to the study of microseismic events during hydraulic fracturing. The spring-slider model idealizes small faults as point sources, reducing the problem to the numerical solution of ordinary differential equations in time. The model provides predictions of moment and moment rate history. The spring-slider idealization provides immense savings in terms of computational expense, as compared to standard geomechanical analysis methods that require spatial discretization of faults or natural fractures that are often vastly smaller than other relevant length scales (e.g., hydraulic fracture length and height, reservoir dimension). Furthermore, with rate-and-state friction, the model predicts if slip is seismic or aseismic.

We applied the spring-slider modeling framework to the simplified problem of a single hydraulic fracture approaching a fault without intersecting it. One-way interaction between the fault and the hydraulic fracture was assumed, and only the leading terms in an asymptotic expansion of the stress field around the hydraulic fracture tip were used in our study. Pore pressure changes and poroelastic effects were also neglected. 

The accuracy of the spring-slider model was assessed through comparison with a finite fault model, solved using finite element analysis in which a fine mesh was placed around the fault to resolve the spatially and temporally variable nucleation and propagation of slip. In general, the spring-slider model was shown to provide accurate predictions of final slip and also the timing of slip (in particular, whether it slip was seismic or aseismic). Quantitative differences emerged when the nucleation length was much smaller than the fault length (i.e., stiffness less than critical stiffness), but these differences might be unimportant for some applications.

With the spring-slider model, we quantified the amount and style (seismic or aseismic) of slip as a function of fault length, orientation, and distance from the hydraulic fracture. Fault slip was only triggered within a narrow range of fault orientations located around the two optimally oriented directions from a Mohr-Coulomb analysis, consistent with the analysis of \citet{zoback2012importance}. 
Furthermore, seismic slip was predicted to occur only within a few meters of the hydraulic fracture, in striking inconsistency to numerous observations of microseismicity occurring several hundred meters from hydraulic fractures \citep{maxwell2002microseismic,rutledge2003hydraulic,shapiro2006hydraulic,maxwell2011enhanced,vermylen2011hydraulic,warpinski2012measurements,friberg2014characterization,duhault2018microseismic,okamoto2018triggering,wilson2018fracking}. This argues for the importance of pore pressure changes from fluid leak-off, changes in elastic stress caused by fracture opening, poroelastic effects, aseismic slip, and possibly other processes, in triggering microseismicity, as has been argued by other authors \citep{bhattacharya2019fluid,eyre2019role,kettlety2019investigating,kettlety2020stress}. 


The spring-slider modeling framework could be extended in a straightforward manner to handle arbitrary loading, including pore pressure changes, and in fact could be embedded in any geomechanics and fluid flow modeling code. The simplest way to do this would be to save time histories of loading stress and pressure at various points (i.e., potential fault locations) in the model, and then one-way couple these into a spring-slider model. The fault lengths and orientations could be drawn from statistical distributions, if there are geostatistical constraints on the natural fracture system in a reservoir, or simply randomly generated. There are many examples of studies employing a stochastic or deterministic fault network \citep[e.g.,][]{ needham1996fault,mcclure2013discrete,mcclure2016fully}.

The one-way coupling assumption neglects interactions between nearby faults. Static stress changes at distance $r$ from a point moment tensor source scale as $M_0/r^3$ \citep{aki2002quantitative}, so one could compare these stress changes to other loading stresses to justify neglecting interactions. Because moment $M_0$ scales with stress drop $\Delta \tau$ and fault length $L$ as $M_0 \sim \Delta \tau L^3$, then static stress changes from small faults decay as $\Delta \tau (L/r)^3$. For typical stress drops of $\Delta \tau \sim 1$ MPa and microseismic fault dimensions of $\sim$1 m, then stress changes drop below 0.1 MPa after only $\sim$10 m. Either larger faults, or successive stress transfer associated with triggered slip across a set of very closely spaced faults, would be necessary to transmit elastic stress changes across large distances.

It is furthermore possible to generalize the spring-slider modeling to account for stress interactions. This leads to a modeling approach that is in many ways quite similar to that taken in earthquake simulators \citep{tullis2012generic} like RSQSim \citep{richards2012rsqsim}, involving multiple fault patches that interact through quasi-static stress transfer.

To close, we have introduced an efficient modeling framework that, with some extensions, could be used to test various hypotheses regarding microseismicity generation during hydraulic fracturing. Once coupled with more sophisticated and realistic descriptions of reservoir stress and pore pressure changes during hydraulic fracturing, this type of model could be used to quantify various aspects of stimulation, such as the relative amount of aseismic to seismic slip.

\section*{Data and Resources}
The spring-slider code is available at
\\
https://github.com/k-rupture/SpringSlider.

\section*{Acknowledgments}
This project was funded by the Stanford Center for Induced and Triggered Seismicity (SCITS). Support from NSF-CMMI-166245 is gratefully acknowledged.

\bibliographystyle{agu04}

\bibliography{main.bbl}

\section*{Appendix}\label{Appendix}
\setcounter{equation}{0}
\setcounter{section}{1}
\renewcommand{\thesection}{\Alph{section}}
\numberwithin{equation}{section}

In this section we discuss the finite element solution of the elasticity problem that provides the mapping between slip $\delta$ and quasi-static shear and normal tractions on the fault, $\Delta \tau^{QS}$ and $\Delta \sigma^{QS}$. As pointed out in the {\bf Introduction}, finite elements have been used by other authors for earthquake modeling studies \citep[e.g., ][]{duan2005multicycle, aagaard2013domain, douilly2017simulation, franceschini2019block, ma2019hybrid, perrin2019persistent} so here we emphasize some key differences in our approach.




We employ a standard Galerkin finite-element approximation to solve the 2-D plane strain linear elasticity equations in a finite domain $\Omega$ (Figure \ref{Fig3}). The solid is homogeneous and isotropic. We use the domain decomposition approach of \citet{aagaard2013domain} to enforce interface conditions on the fault $\Gamma_F$, specifically imposed slip,
\begin{equation}
(\textit{\textbf{u}}^+ - \textit{\textbf{u}}^-)\cdot\textit{\textbf{s}}= \delta \textrm{ on }  \Gamma_F,
\label{Eq15}
\end{equation}
no opening,
\begin{equation}
(\textit{\textbf{u}}^+ - \textit{\textbf{u}}^-)\cdot\textit{\textbf{n}}= 0 \textrm{ on }  \Gamma_F,
\label{Eq16}
\end{equation}
and equal and opposite tractions on opposing sides of the fault
\begin{equation}
\textit{\textbf{t}}^+ + \textit{\textbf{t}}^-= \textbf{0} \textrm{ on }  \Gamma_F.
\label{Eq17}
\end{equation}
The displacement vector is $\textit{\textbf{u}}$ and superscripts $\pm$ denote values on the two sides of the fault. The unit normal and tangent vectors to the fault are $\textit{\textbf{n}}$ and $\textit{\textbf{s}}$, respectively, with $\textit{\textbf{n}}$ pointing from the negative to positive side of the fault. The traction vectors on the positive and negative sides of the fault are $\textit{\textbf{t}}^{\pm}$. In all examples in this study, we enforce zero displacement boundary conditions on the external boundaries of the computational domain, which are placed sufficiently far from the fault that this has negligible influence on the slip process.

The finite element method adopted to approximate the elasticity equation is based on a standard formulation for the field equations and Dirichlet and Neumann boundary conditions \citep[see classical finite element books e.g.,][]{reddy1993introduction,zienkiewicz2005finite}. The kinematic constraints enforcing the displacement discontinuity (equations (\ref{Eq15}) and (\ref{Eq16})) are imposed through the Lagrange multipliers along the fault. As a result, the continuity of the tractions (equation (\ref{Eq17})) follows in an “averaged” way. We use linear elements for both the displacement field in the domain and the Lagrange multipliers along the fault.

In contrast to \citet{aagaard2013domain}, we do not use the Lagrange multipliers as approximations for the tractions when solving the rate-and-state equations. This is because the Lagrange multiplier tractions suffer from oscillations near the fault tips \citep[e.g.,][]{liu2010stabilized}.  


Instead of evaluating the tractions from the Lagrange multipliers, we utilize a stress-smoothing technique that has been described in the literature \citep[e.g., ][]{oden1971calculation, hinton1974local,mijuca1997stress,carrera2001developments} and has been used by the first author \citep[see Section 3.3 of][]{kashefi2018finite} to obtain the shear stress from the velocity space for fluid dynamics applications. However, to the best of our knowledge, this technique has not been used before in fault or earthquake modeling. For our problem, this approach rendered good results.

We explain the stress-smoothing technique in terms of $\mathbb{P}_1$ elements for displacement and 2D plane strain elasticity; however, the approach is applicable to other elements and to 3D. The obvious alternative to the values of the Lagrange multipliers would be to evaluate the tractions on the fault from the computed stress in the finite element solution. However, two issues suggest the need for an alternative: 1.) the resulting stress field is discontinuous at element boundaries, and hence does not have a unique value at the nodes on the fault where the rate-and-state equation (equation (\ref{Eq32})) is integrated, and 2.) such stress field is not guaranteed to converge in the space of square integrable functions along the fault, and much less pointwise.     


Instead, we proceed by performing an $L_2$-projection of the computed stress field, which is constant in each element, over the space of continuous piecewise linear ($\mathbb{P}_1$) elements in the domain. By evaluating the resulting stress field along the fault, a continuous traction field follows. In this way, this approach solves the first aforementioned issue. We cannot guarantee pointwise convergence for this field either, but the consistent observation across all of our numerical experiments is that the averaging effect of the projection renders very smooth traction fields along the fault, returning what seem reasonable results. 

To specify this method, we first write the components of the stress tensor in 2D plane strain: 

\begin{equation}
\sigma_{xx}=\frac{2G}{1-\nu}\frac{\partial u_x}{\partial x} + \frac{2G\nu}{1-\nu}\frac{\partial u_y}{\partial y},
\label{EqA1}
\end{equation}

\begin{equation}
\sigma_{xy}=G\left(\frac{\partial u_x}{\partial y} + \frac{\partial u_y}{\partial x}\right),
\label{EqA2}
\end{equation}

\begin{equation}
\sigma_{yy}=\frac{2G\nu}{1-\nu}\frac{\partial u_x}{\partial x}+\frac{2G}{1-\nu}\frac{\partial u_y}{\partial y}.
\label{EqA3}
\end{equation}
Next, we consider the weak form of equations (\ref{EqA1})-(\ref{EqA3}) and approximate the space of stress and displacement fields using $\mathbb{P}_1$ elements as well as $\mathbb{P}_1$ trial functions. The resulting matrix form is

\begin{equation}
\textrm{\textbf{M}}P=\textrm{\textbf{B}}U,
\label{EqA4}
\end{equation}
where \textbf{M} is the mass matrix, \textbf{B} is the gradient matrix associated with equations (\ref{EqA1})-(\ref{EqA3}), and \textit{P} is a vector containing the nodal values of stress tensor components computed by the stress-smoothing technique. Solution of equation (\ref{EqA4}) provides stresses at all nodes in the domain. From these, we compute the traction vector at all fault nodes. 

Next we provide examples that demonstrate the performance of this stress-smoothing technique as compared to using the Lagrange multipliers directly. We consider a fault embedded in a computational domain (Figure \ref{Fig3}b) with zero-displacement boundary conditions on the external sides of the domain. We impose a slip distribution along the fault and compute the displacement field. Then we apply the stress-smoothing technique by solving equation (\ref{EqA4}) and resolve these stresses into tractions on the fault. Figures \ref{Fig14}b and \ref{Fig14}d show the results for elliptical and sinusoidal slip distributions. Tractions from the stress-smoothing technique do not have the oscillatory behaviour of the Lagrange multipliers near the fault tips, even when a stress singularity exists at the fault tips.

To evaluate the method's performance, we conducted a convergence test. Figure \ref{Fig15}a shows  shear traction on the fault, obtained with the stress-smoothing technique at multiple levels of resolution for an imposed elliptical distribution of slip. While the exact solution is a constant shear traction on the fault, due to the singularity of the stress field on the "other sides" of the fault tips, a spurious singularity appears on the shear tractions on the fault. The width  of the singularity on the fault shrinks as the mesh size is reduced, and very accurate results are obtained away from the fault tips.  In Fig. \ref{Fig15}b, we plot the norms of the error with respect to the exact solution \citep[for the exact solution see e.g.,][]{starr1928slip}. In agreement with the aforementioned observation, the mean-square error ($L_2$) was computed on the entire fault, the interval $[4.5, 5.5]$, showing a lack of convergence. Instead,  the mean-square error on 80\% of the fault, the interval $[4.6, 5.4]$, converges,  decreasing linearly with the number of element on the fault. This follows the same expected rate than the error in the  stresses in the entire domain (not shown), namely, with the square root of the largest edge length of a triangle in the mesh. Additionally, we plotted the  maximum error ($L_\infty$) over the same $80\%$ of the fault, observing a similar trend. This result is reassuring given that the integration of the rate-and-state friction involves the pointwise evaluation of the tractions on the fault. 
 
The results in this example are encouraging, but a more detailed analysis is needed to learn when such benign behavior 
could be expected.

\section*{Full Mailing Address for Each Author}
Ali Kashefi\\
Department of Mechanical Engineering, Stanford University\\
440 Escondido Mall\\
Stanford, CA 94305, USA\\

Eric M. Dunham\\
Department of Geophysics, Stanford University\\
397 Panama Mall\\
Stanford, CA 94305, USA\\

Benjamin Grossman-Ponemon\\
Department of Mechanical Engineering, Stanford University\\
440 Escondido Mall\\
Stanford, CA 94305, USA\\

Adrian J. Lew\\
Department of Mechanical Engineering, Stanford University\\
440 Escondido Mall\\
Stanford, CA 94305, USA\\

\section*{List of Figure Captions}

\begin{figure}
\centering
\includegraphics[width=1.0\textwidth]{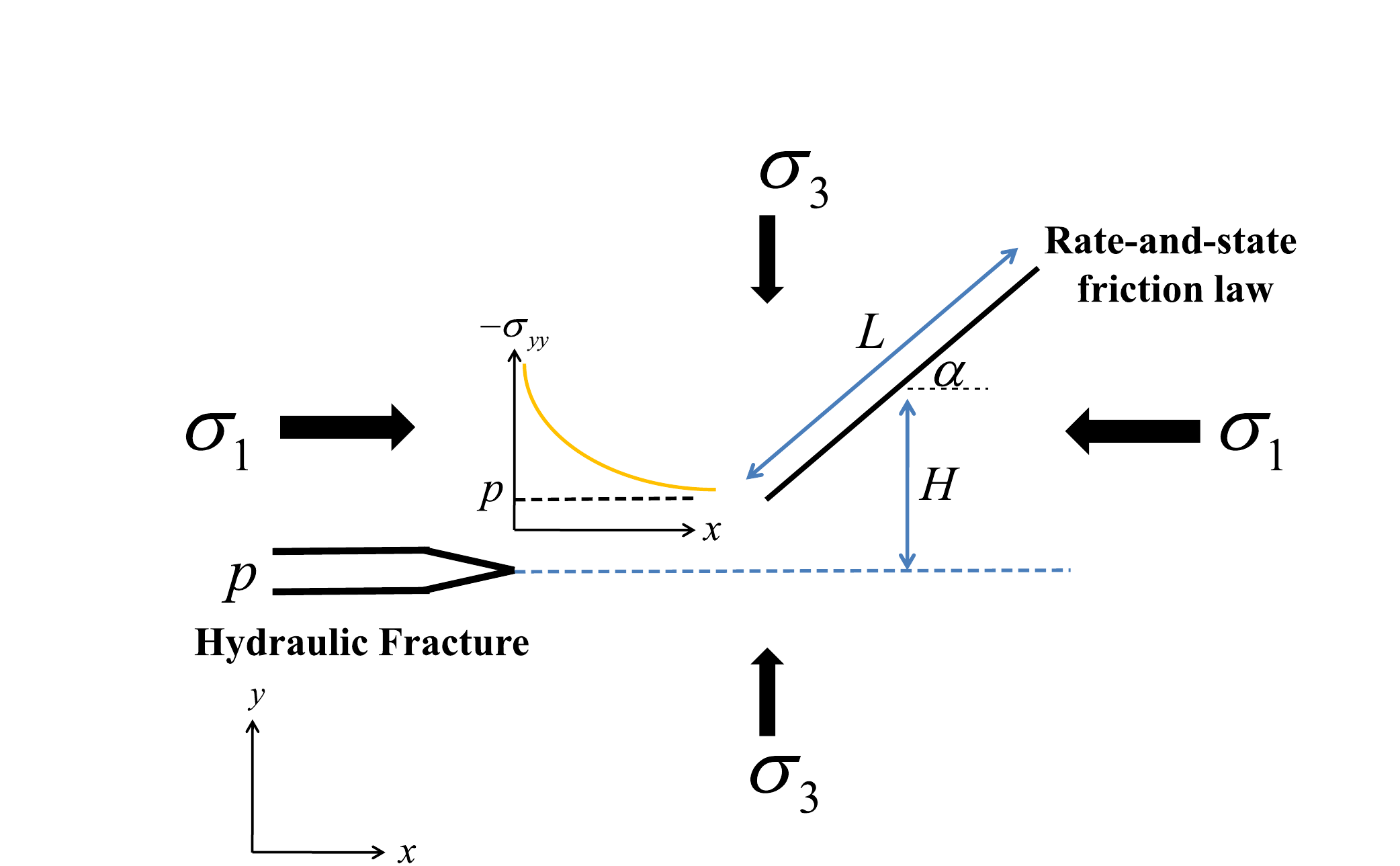}
\caption[C]{Schematic representation of a fault of length of $L$ and orientation $\alpha$, and a hydraulic fracture with fluid pressure $p$; the hydraulic fracture moves on a straight line that passes at distance $H$ from the fault center.}
\label{Fig1}
\end{figure}

Figure \ref{Fig1}: Schematic representation of a fault of length of $L$ and orientation $\alpha$, and a hydraulic fracture with fluid pressure $p$; the hydraulic fracture moves on a straight line that passes at distance $H$ from the fault center.

\begin{figure}
\centering
\includegraphics[width=1.0\textwidth]{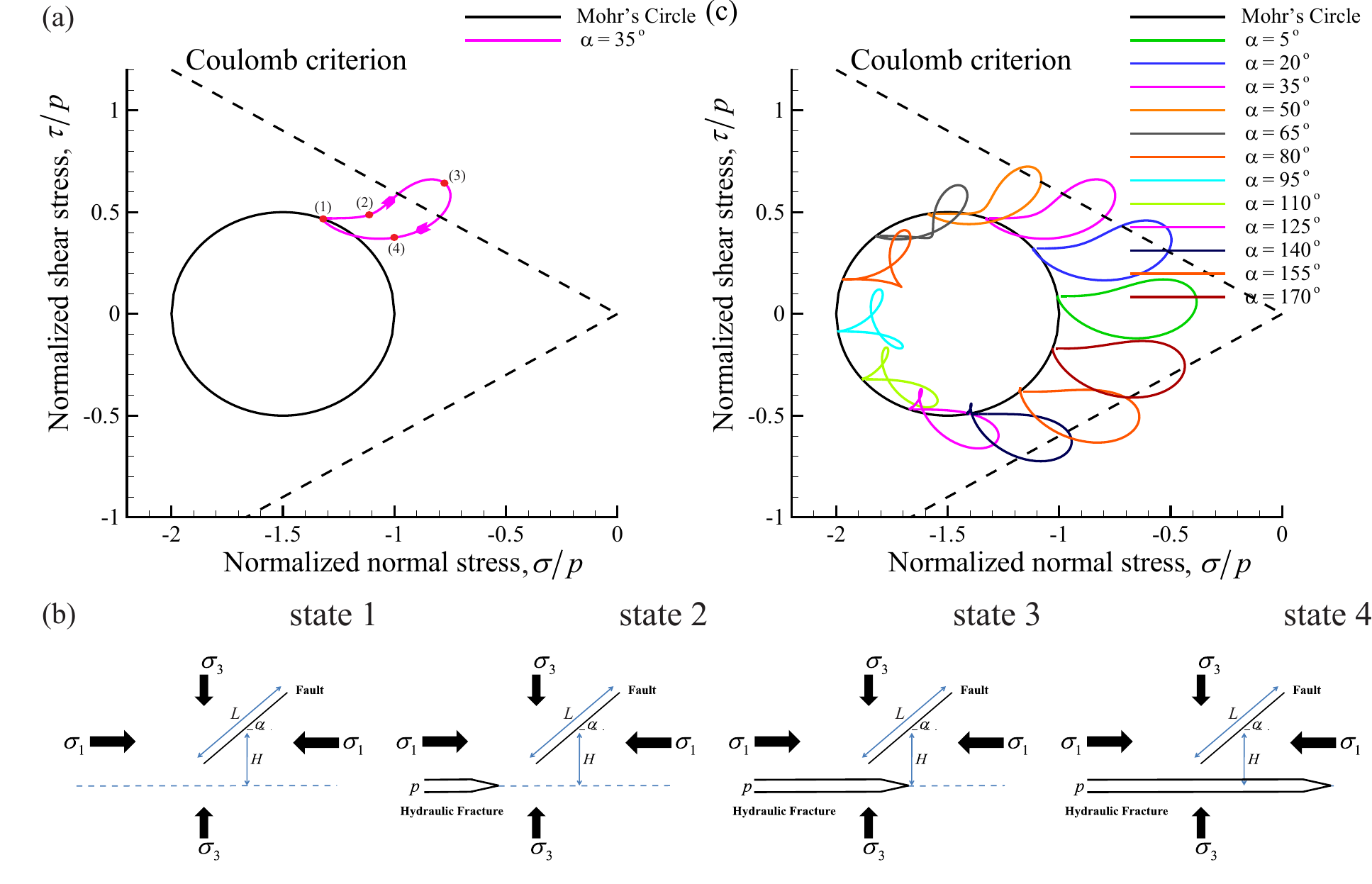}
\caption[C]{(a) Mohr circle for spatially uniform stress state (black) and trajectory (pink) showing stress evolution on a fault oriented at $\alpha=35^{\circ}$ from the hydraulic fracture; (b) Stress trajectory for faults with various orientations; (c) Hydraulic fracture position with respect to the fault for four different states shown in (a).}
\label{Fig2}
\end{figure}

Figure \ref{Fig2}: (a) Mohr circle for spatially uniform stress state (black) and trajectory (pink) showing stress evolution on a fault oriented at $\alpha=35^{\circ}$ from the hydraulic fracture; (b) Stress trajectory for faults with various orientations; (c) Hydraulic fracture position with respect to the fault for four different states shown in (a).

\begin{figure}
\centering
\includegraphics[width=1.00\textwidth]{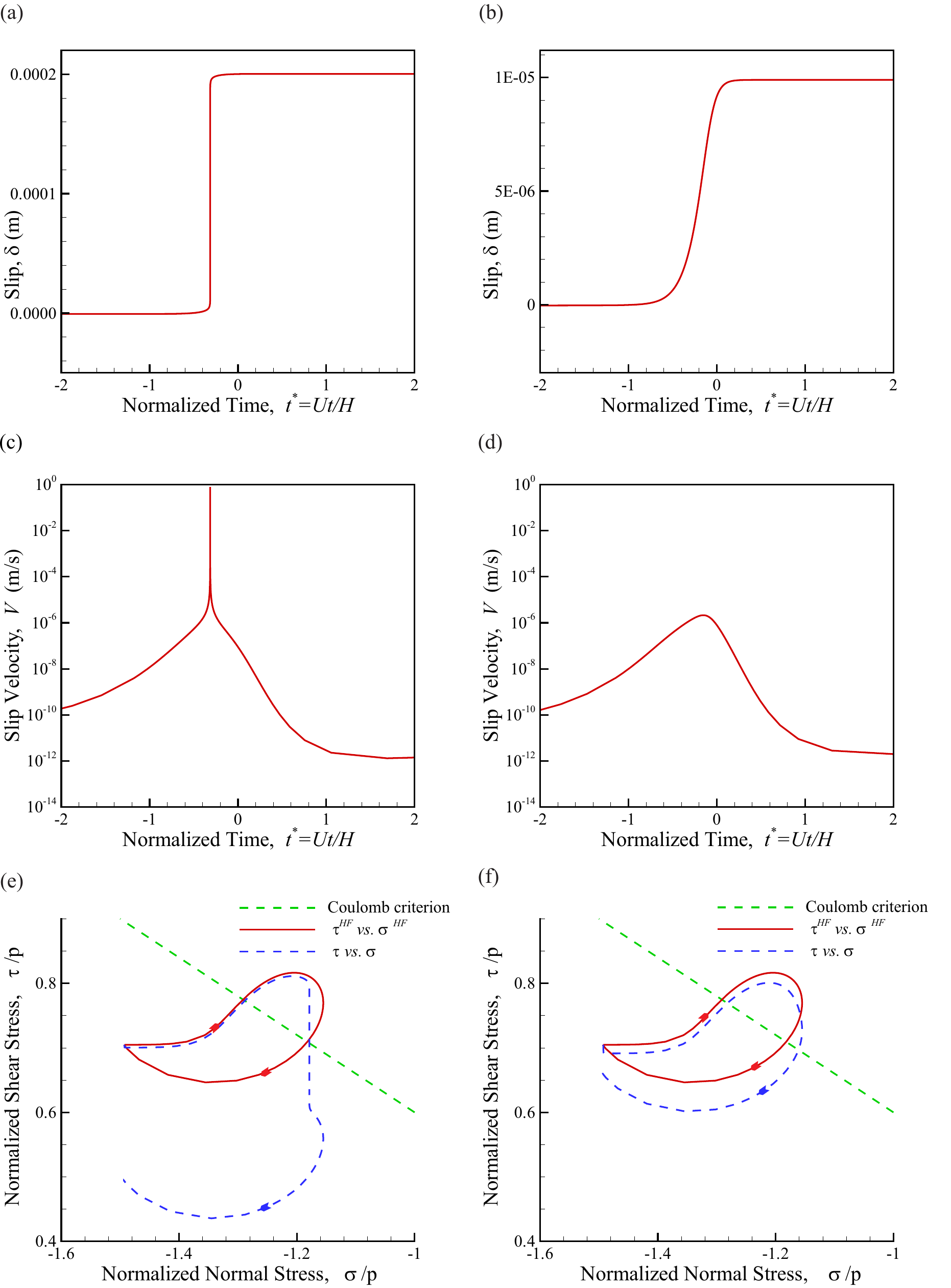}
\caption[C]{Spring-slider solutions for $L=1.3$ m (seismic slip, left column) and $L=0.4$ m (aseismic slip, right column), showing (a)-(b) slip, (c)-(d) slip velocity, and (e)-(f) stress.}
\label{Fig4}
\end{figure}

Figure \ref{Fig4}: Spring-slider solutions for $L=1.3$ m (seismic slip, left column) and $L=0.4$ m (aseismic slip, right column), showing (a)-(b) slip, (c)-(d) slip velocity, and (e)-(f) stress.

\begin{figure}
\centering
\includegraphics[width=1.00\textwidth]{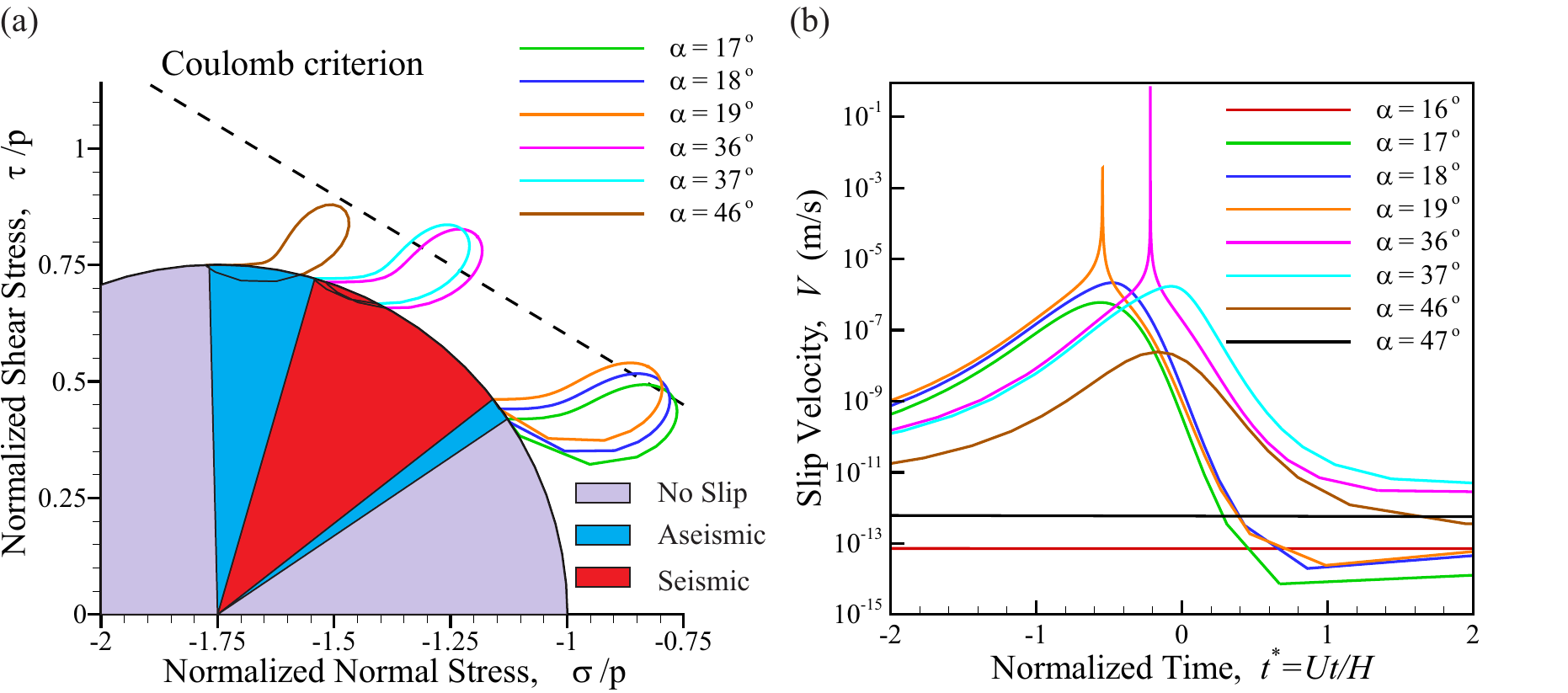}
\caption[C]{Influence of fault orientation on sliding style. (a) Stress trajectory for different fault orientations, with slip style (seismic, aseismic, no slip) marked in color. (b) Evolution of slip velocity. Seismic slip is observed for $37^{\circ} \le \alpha \le 46^{\circ}$. As orientation moves away from the seismic orientations, there is a transition to aseismic slip that diminishes eventually to negligible values.}
\label{Fig5}
\end{figure}

Figure \ref{Fig5}: Influence of fault orientation on sliding style. (a) Stress trajectory for different fault orientations, with slip style (seismic, aseismic, no slip) marked in color. (b) Evolution of slip velocity. Seismic slip is observed for $37^{\circ} \le \alpha \le 46^{\circ}$. As orientation moves away from the seismic orientations, there is a transition to aseismic slip that diminishes eventually to negligible values. 

\begin{figure}
\centering
\includegraphics[width=1.00\textwidth]{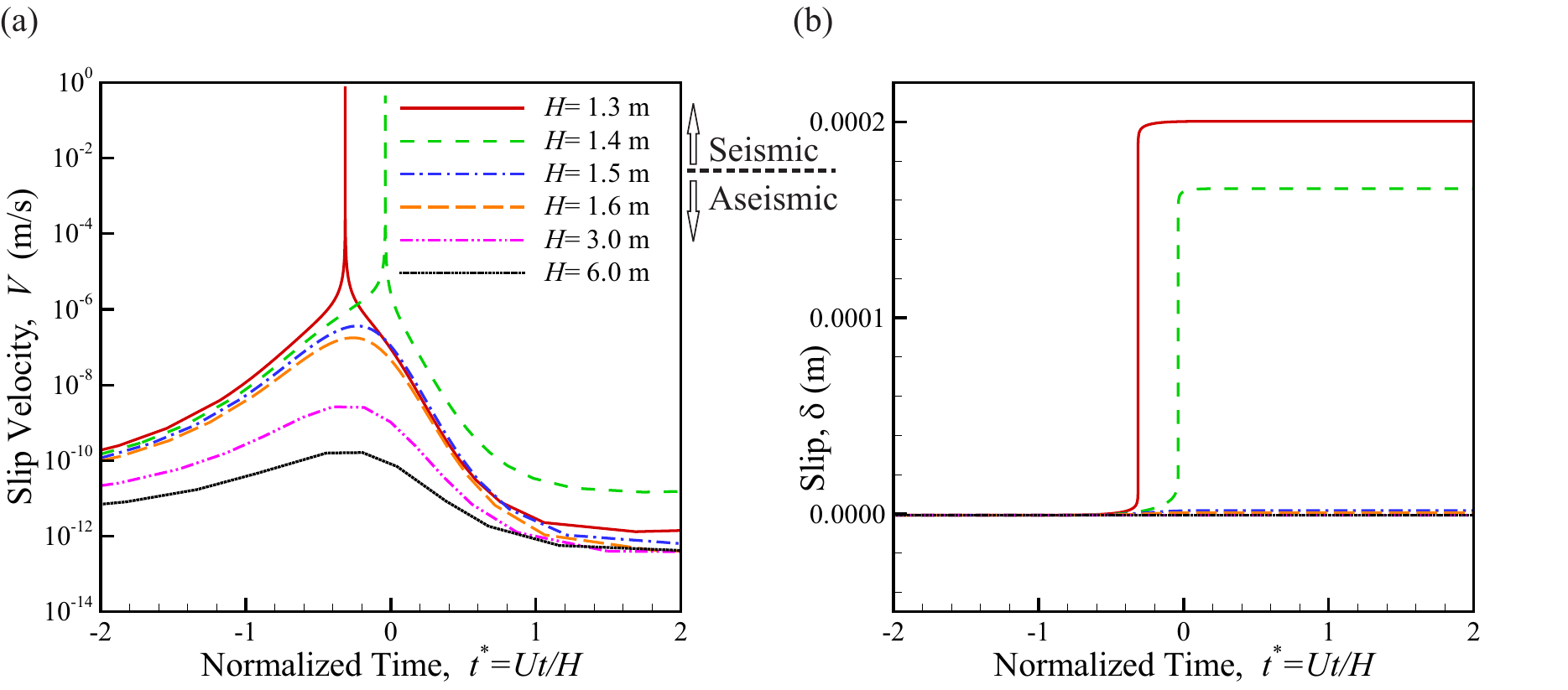}
\caption[C]{(a) Slip velocity and (b) slip from a spring-slider exploration of the effect of the distance $H$ between the hydraulic fracture and the fault center.}
\label{Fig6}
\end{figure}

Figure \ref{Fig6}: (a) Slip velocity and (b) slip from a spring-slider exploration of the effect of the distance $H$ between the hydraulic fracture and the fault center.

\begin{figure}
\centering
\includegraphics[width=1.00\textwidth]{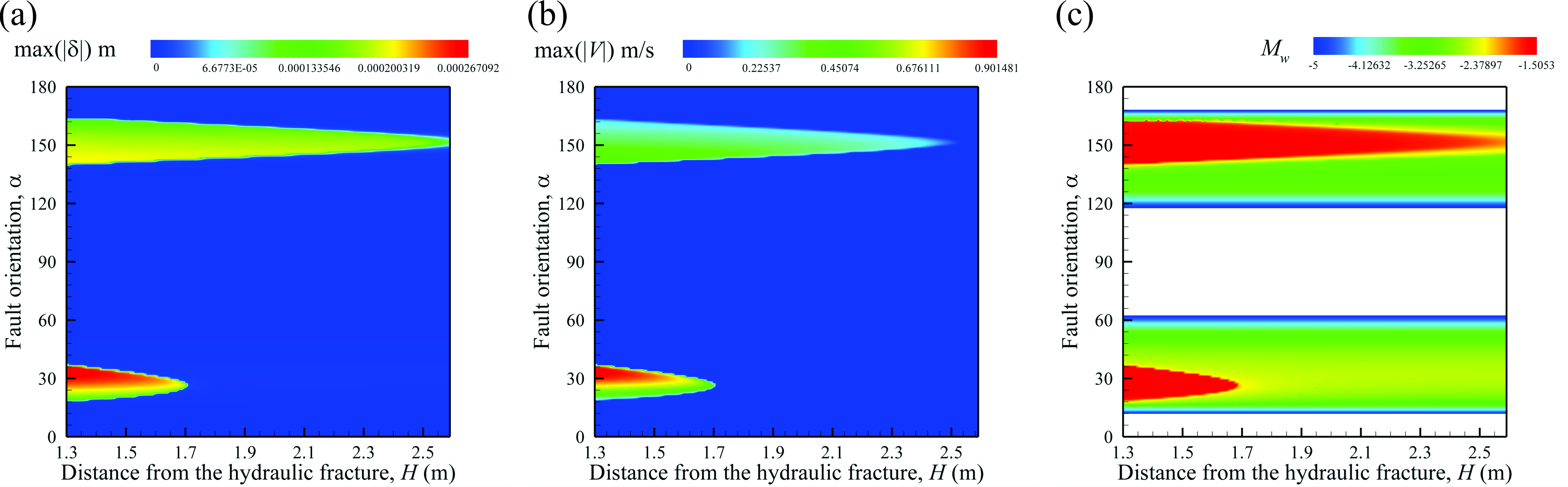}
\caption[C]{Contours of (a) maximum absolute value of slip, (b) maximum absolute value of slip velocity, and (c) magnitude $M_w$ as a function of $H$ and $\alpha$.}
\label{Fig7}
\end{figure}

Figure \ref{Fig7}: Contours of (a) maximum absolute value of slip, (b) maximum absolute value of slip velocity, and (c) magnitude $M_w$ as a function of $H$ and $\alpha$.

\begin{figure}
\centering
\includegraphics[width=1.00\textwidth]{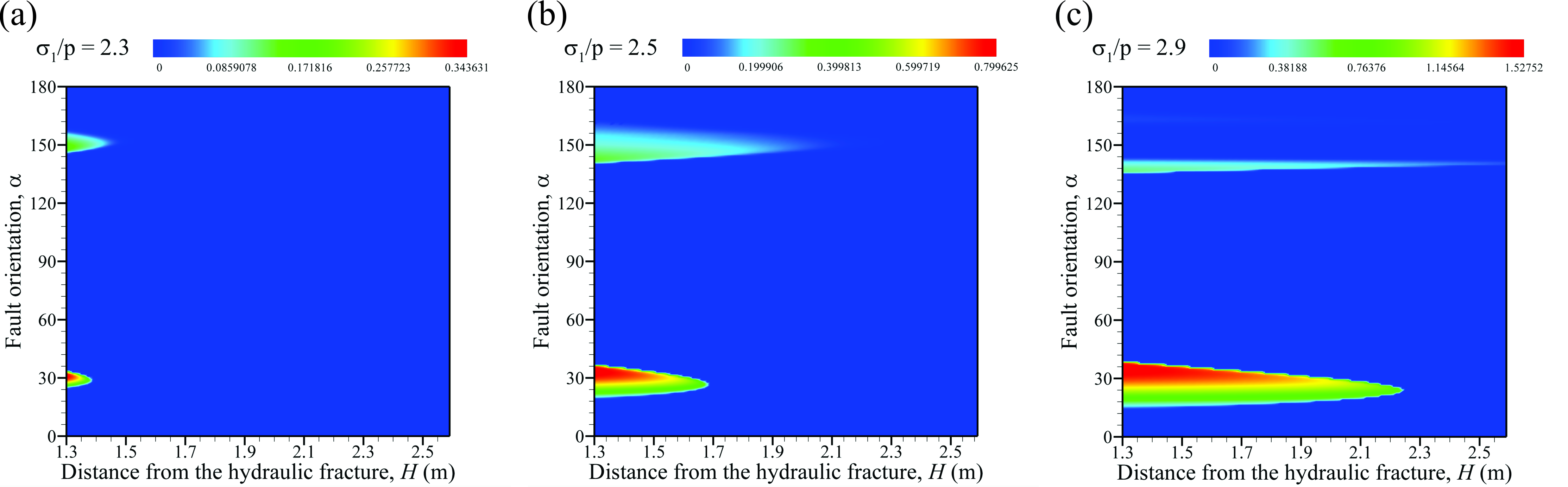}
\caption[C]{Contours of maximum absolute value of slip velocity for (a) $\sigma_1/p=2.3$, (b) $\sigma_1/p=2.5$, and (c) $\sigma_1/p=2.9$ as a function of $H$ and $\alpha$.}
\label{Fig7_1}
\end{figure}

Figure \ref{Fig7_1}: Contours of maximum absolute value of slip velocity for (a) $\sigma_1/p=2.3$, (b) $\sigma_1/p=2.5$, and (c) $\sigma_1/p=2.9$ as a function of $H$ and $\alpha$.

\begin{figure}
\centering
\includegraphics[width=1.00\textwidth]{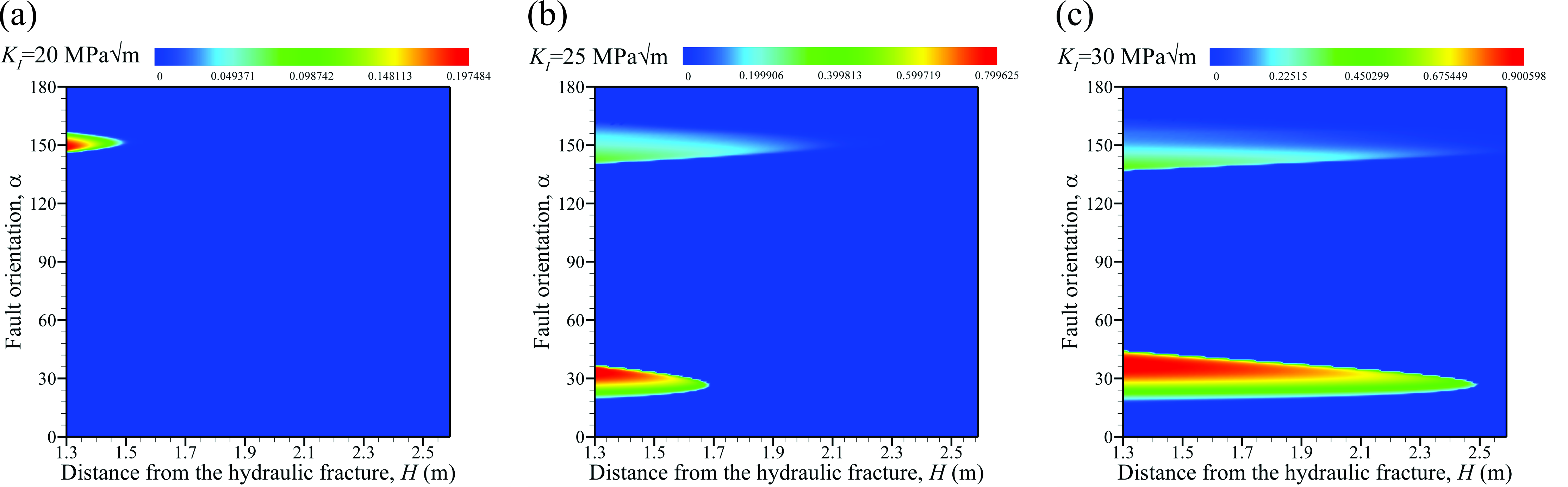}
\caption[C]{Contours of maximum absolute value of slip velocity for (a) $K_I=20$ MPa$\sqrt{\textrm{m}}$, (b) $K_I=25$ MPa$\sqrt{\textrm{m}}$, and (c) $K_I=30$ MPa$\sqrt{\textrm{m}}$ as a function of $H$ and $\alpha$.}
\label{Fig7_2}
\end{figure}

Figure \ref{Fig7_2}: Contours of maximum absolute value of slip velocity for (a) $K_I=20$ MPa$\sqrt{\textrm{m}}$, (b) $K_I=25$ MPa$\sqrt{\textrm{m}}$, and (c) $K_I=30$ MPa$\sqrt{\textrm{m}}$ as a function of $H$ and $\alpha$. 

\begin{figure}
\centering
\includegraphics[width=1.00\textwidth]{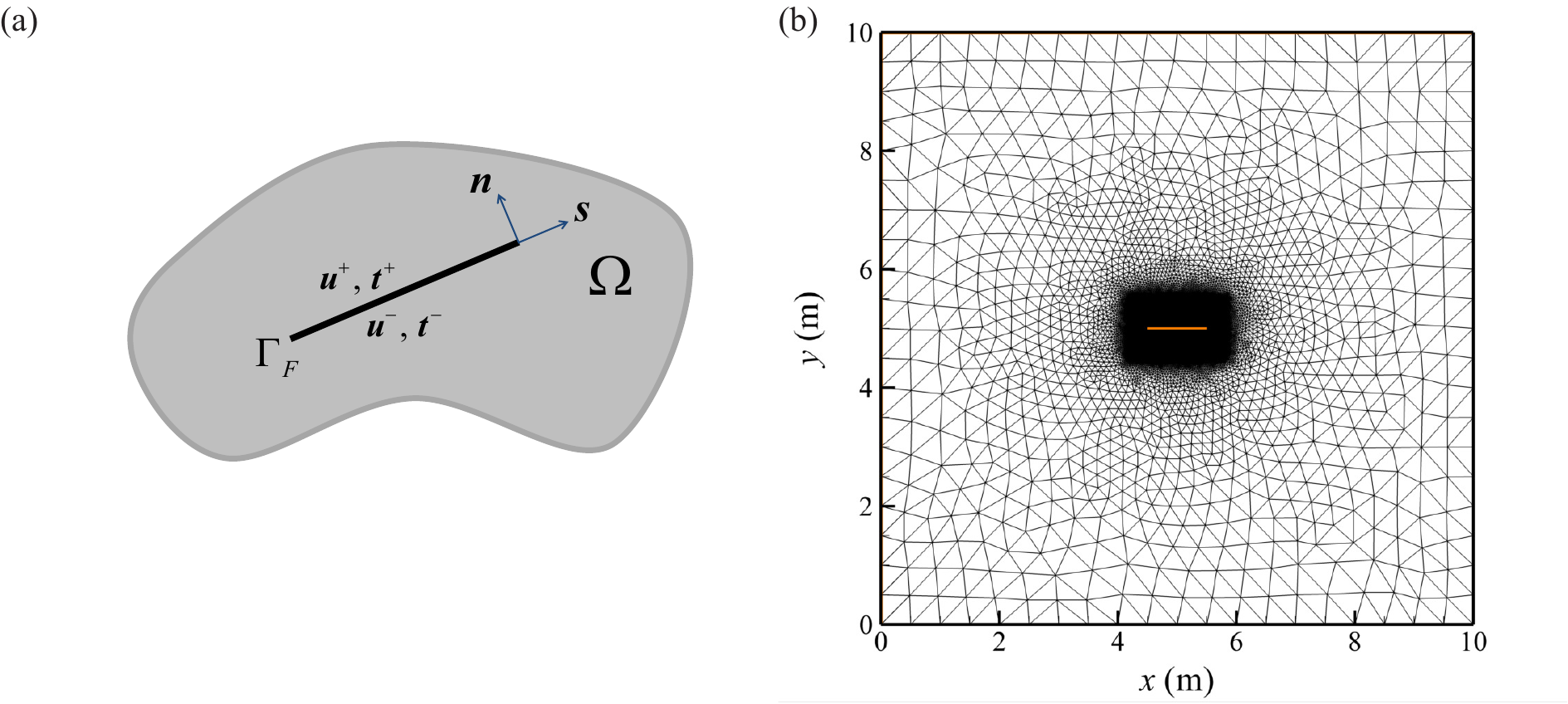}
\caption[C]{(a) Domain decomposition approach for a fault embedded in an elastic solid; the displacement field is discontinuous across the fault while tractions on opposing sides are equal and opposite. (b) Triangular finite element mesh with 22910 nodes and 45416 elements.}
\label{Fig3}
\end{figure}

Figure \ref{Fig3}: (a) Domain decomposition approach for a fault embedded in an elastic solid; the displacement field is discontinuous across the fault while tractions on opposing sides are equal and opposite. (b) Triangular finite element mesh with 22910 nodes and 45416 elements.

\begin{figure}
\centering
\includegraphics[width=1.00\textwidth]{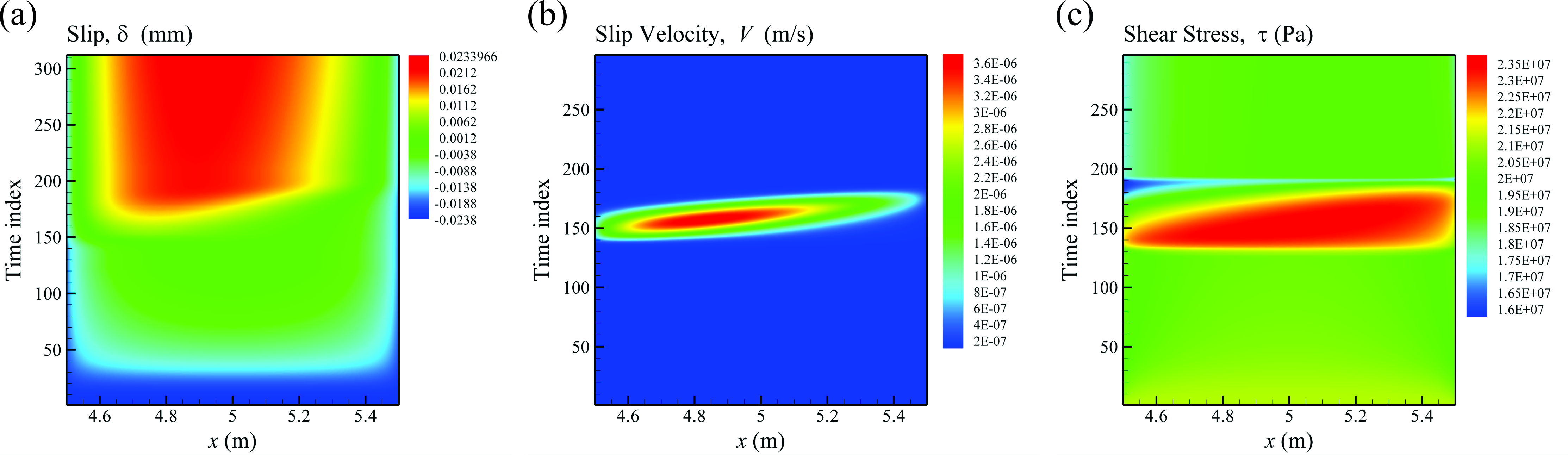}
\caption[C]{Finite element solution for (a) slip, (b) slip velocity, and (c) shear stress along a fault with $d_c=14\times10^{-5}$ m ($k>k_{cr}$).}
\label{Fig8}
\end{figure}

Figure \ref{Fig8}: Finite element solution for (a) slip, (b) slip velocity, and (c) shear stress along a fault with $d_c=14\times10^{-5}$ m ($k>k_{cr}$).    

\begin{figure}
\centering
\includegraphics[width=1.00\textwidth]{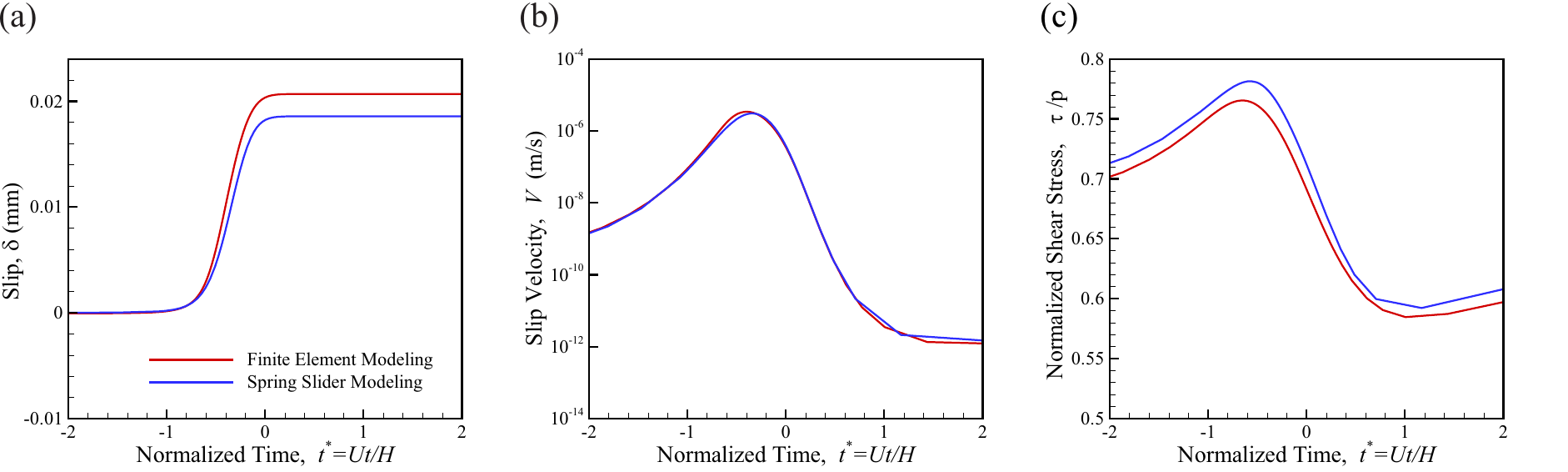}
\caption[C]{Comparison between spring-slider and finite element solutions for (a) slip, (b) slip velocity, and (c) shear stress with $d_c=14\times10^{-5}$ m ($k>k_{cr}$). For the finite element solution, slip and slip velocity are from the center point of the fault, and shear stress is averaged over the fault.}
\label{Fig9}
\end{figure}

Figure \ref{Fig9}: Comparison between spring-slider and finite element solutions for (a) slip, (b) slip velocity, and (c) shear stress with $d_c=14\times10^{-5}$ m ($k>k_{cr}$). For the finite element solution, slip and slip velocity are from the center point of the fault, and shear stress is averaged over the fault. 

\begin{figure}
\centering
\includegraphics[width=1.00\textwidth]{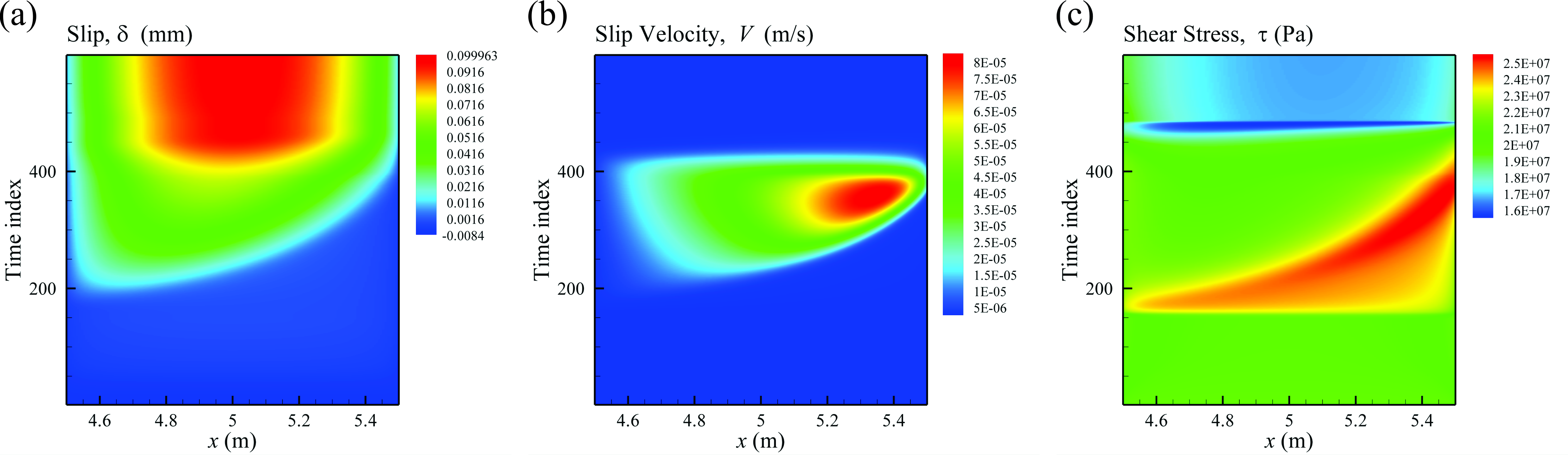}
\caption[C]{Same as Figure \ref{Fig8} but for $d_c=2.4\times10^{-5}$ m ($k\approx k_{cr}$).}
\label{Fig10}
\end{figure}

Figure \ref{Fig10}: Same as Figure \ref{Fig8} but for $d_c=2.4\times10^{-5}$ m ($k\approx k_{cr}$). 

\begin{figure}
\centering
\includegraphics[width=1.00\textwidth]{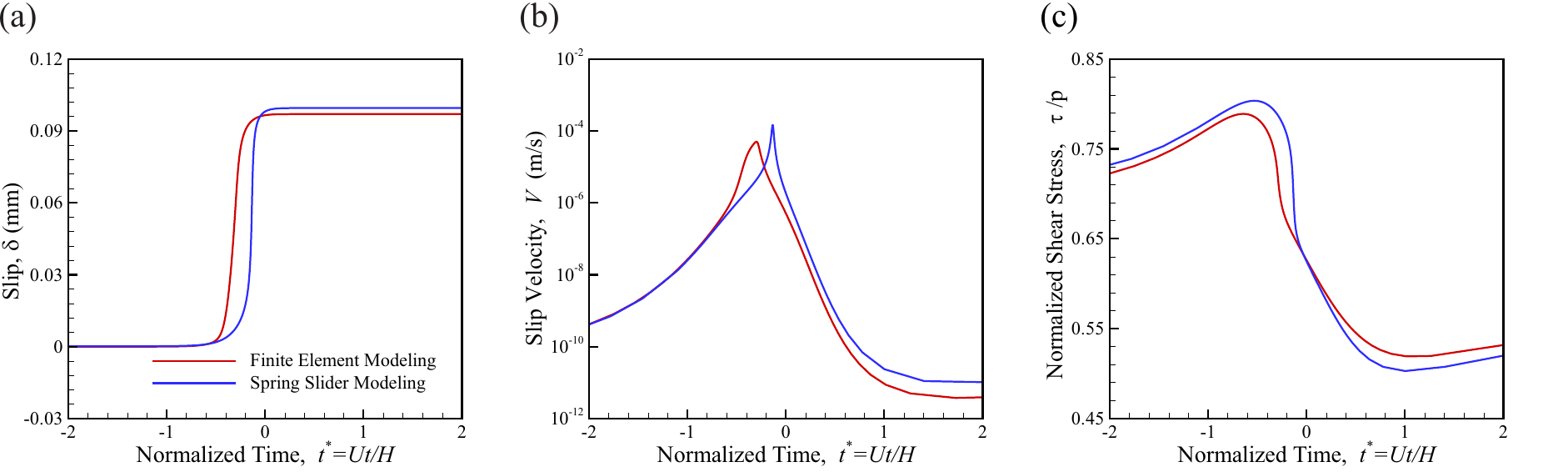}
\caption[C]{Same as Figure \ref{Fig9} but for $d_c=2.4\times10^{-5}$ m ($k\approx k_{cr}$).}
\label{Fig11}
\end{figure}

Figure \ref{Fig11}: Same as Figure \ref{Fig9} but for $d_c=2.4\times10^{-5}$ m ($k\approx k_{cr}$).

\begin{figure}
\centering
\includegraphics[width=1.00\textwidth]{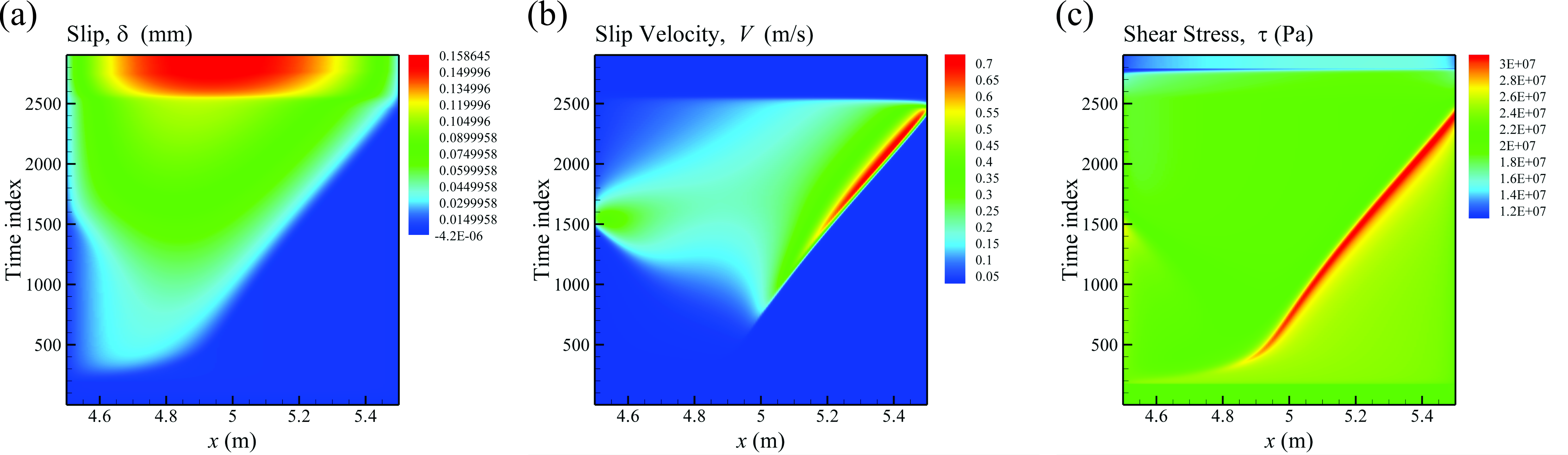}
\caption[C]{Same as Figure \ref{Fig8} but for $d_c=1\times10^{-5}$ m ($k<k_{cr}$).}
\label{Fig12}
\end{figure}

Figure \ref{Fig12}: Same as Figure \ref{Fig8} but for $d_c=1\times10^{-5}$ m ($k<k_{cr}$).

\begin{figure}
\centering
\includegraphics[width=1.00\textwidth]{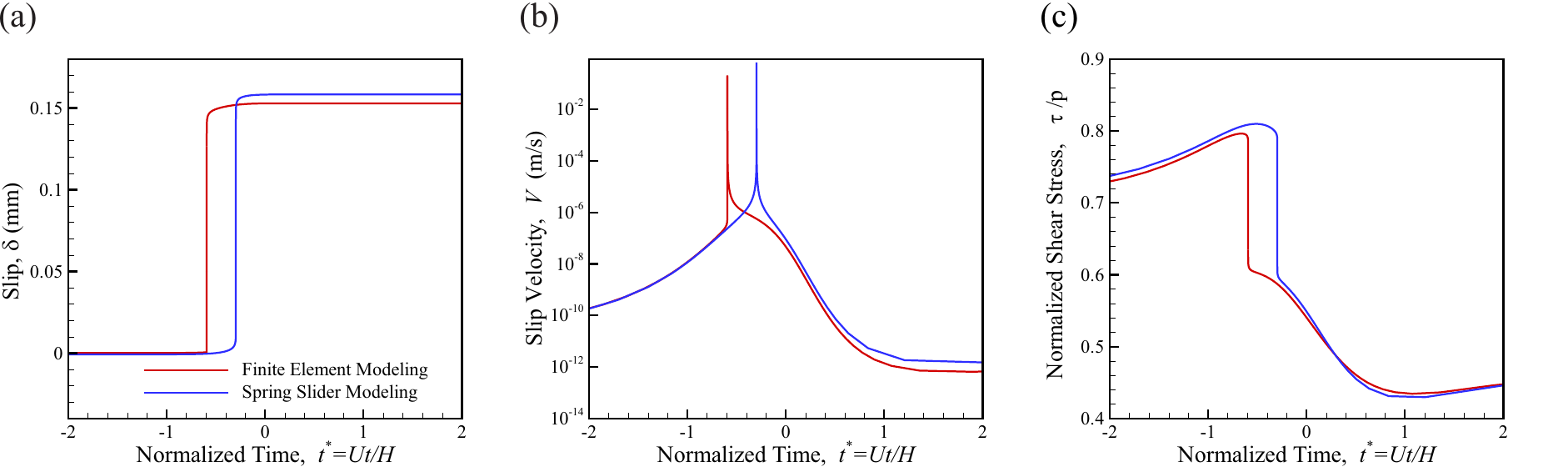}
\caption[C]{Same as Figure \ref{Fig9} but for $d_c=1\times10^{-5}$ m ($k<k_{cr}$).}
\label{Fig13}
\end{figure}

Figure \ref{Fig13}: Same as Figure \ref{Fig9} but for $d_c=1\times10^{-5}$ m ($k<k_{cr}$).

\begin{figure}
\centering
\includegraphics[width=1.00\textwidth]{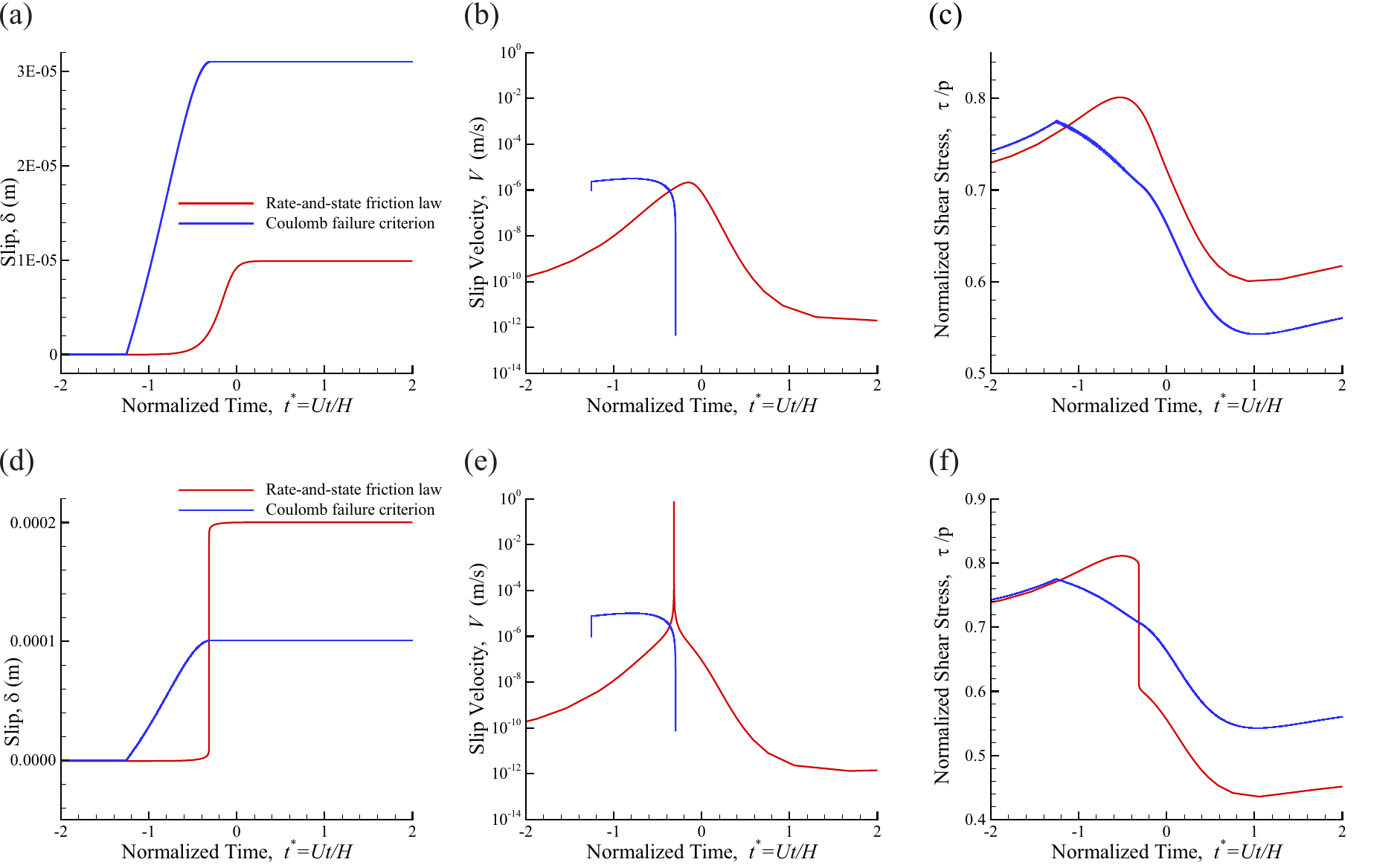}
\caption[C]{Comparison of rate-and-state friction and Coulomb failure criterion for (a)-(c) $L=0.4$ m and (d)-(f) $L=1.3$ m.}
\label{Fig13_1}
\end{figure}

Figure \ref{Fig13_1}: Comparison of rate-and-state friction and Coulomb failure criterion for (a)-(c) $L=0.4$ m and (d)-(f) $L=1.3$ m. 

\begin{figure}
\centering
\includegraphics[width=1.00\textwidth]{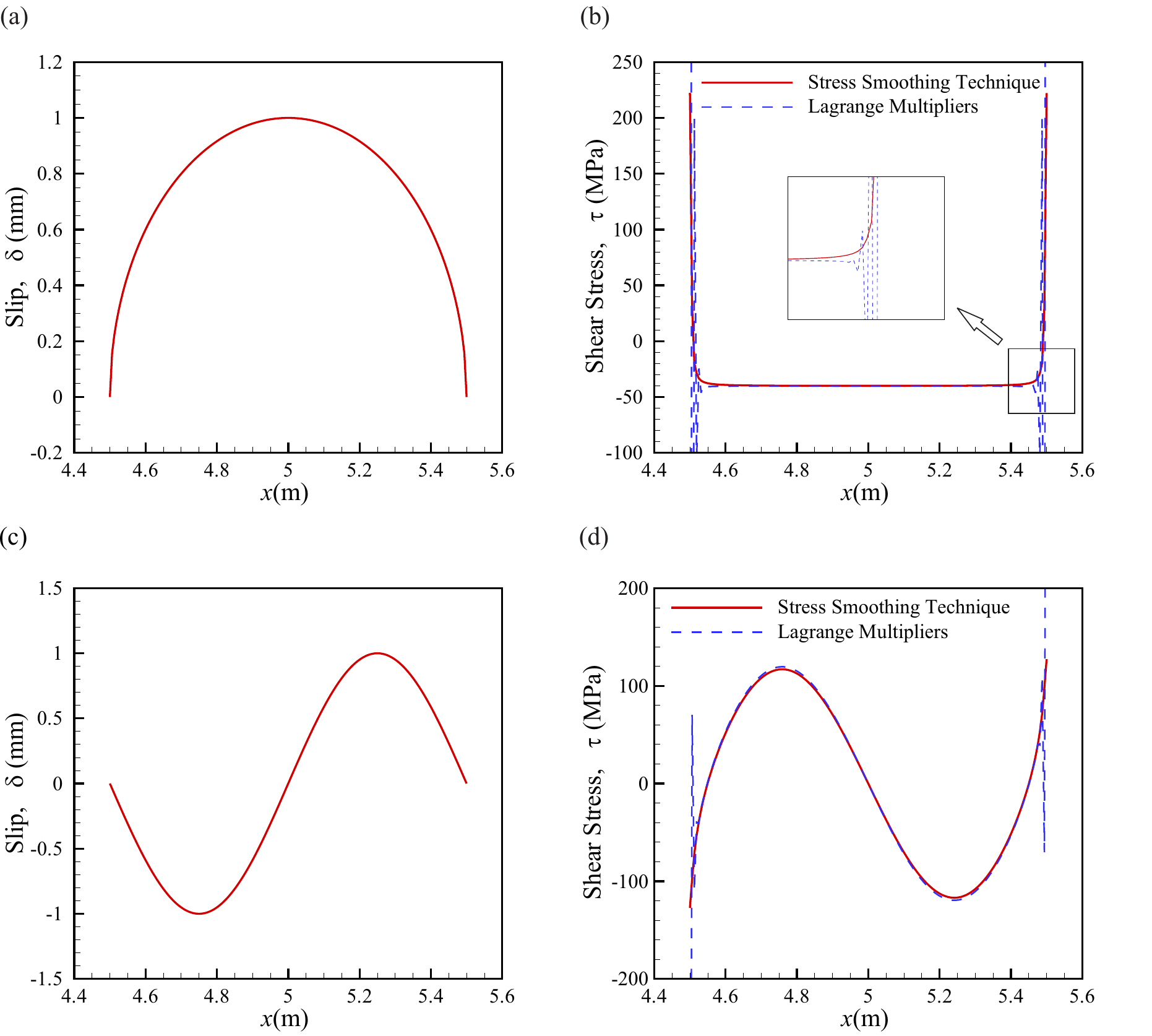}
\caption[C]{Comparison of Lagrange multipliers and stress-smoothing technique. (a) Elliptical slip distribution imposed on the fault, and (b) associated shear stress change on the fault. Insets (c) and (d) show the same for a  sinusoidal slip distribution.}
\label{Fig14}
\end{figure}

Figure \ref{Fig14}: Comparison of Lagrange multipliers and stress-smoothing technique. (a) Elliptical slip distribution imposed on the fault, and (b) associated shear stress change on the fault. (c) and (d) Same for sinusoidal slip distribution.

\begin{figure}
\centering
\includegraphics[width=1.00\textwidth]{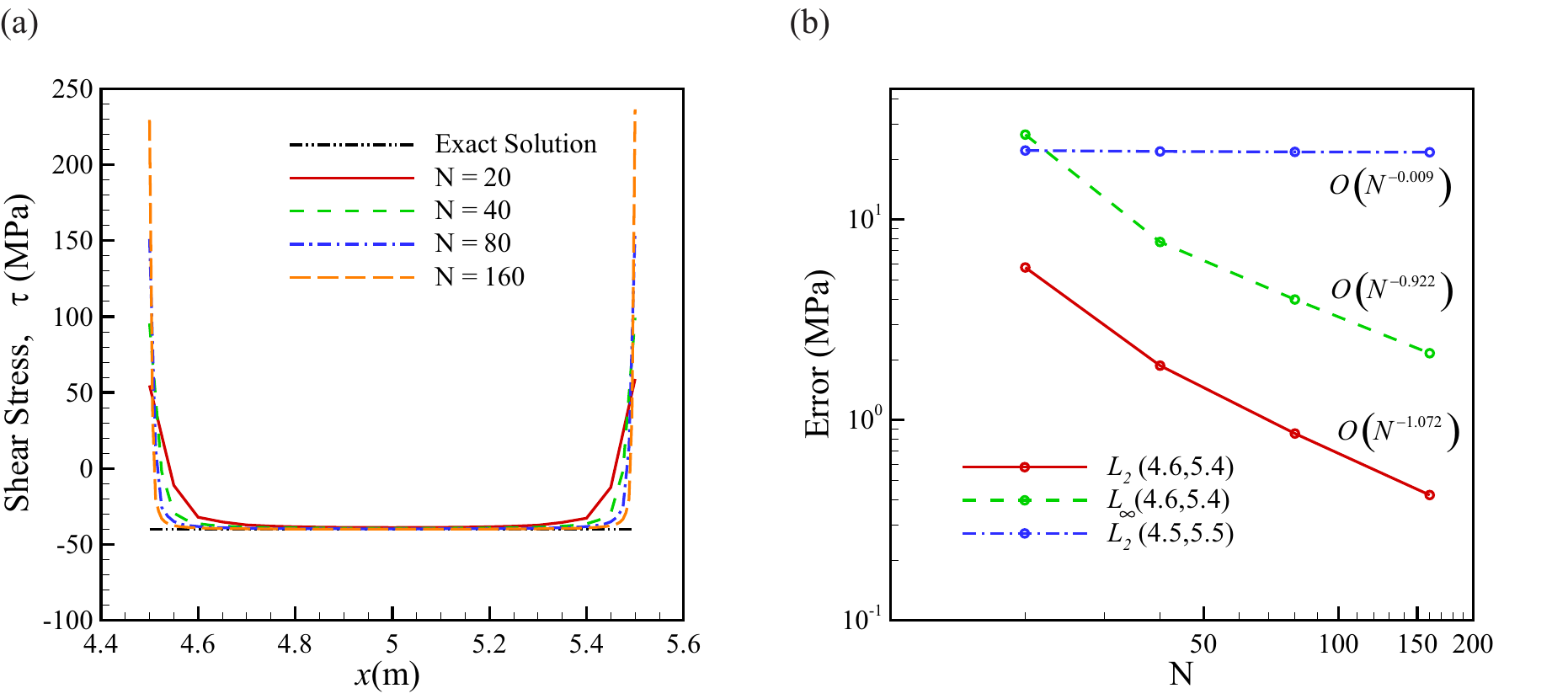}
\caption[C]{Convergence test for the stress-smoothing technique. (a) Shear stress on the fault at multiple resolutions for an imposed elliptical slip distribution and the exact solution. (b) Error of the numerical solution as a function of the number $N$ of elements along the fault, computed over the entire fault ([4.5,5,5]) and on 80\% of the fault ([4.6,5.4]). Convergence is observed in the interior of the fault, but not near the fault tips.}
\label{Fig15}
\end{figure}

Figure \ref{Fig15}: Convergence test for the stress-smoothing technique. (a) Shear stress on the fault at multiple resolutions for an imposed elliptical slip distribution and the exact solution. (b) Error of the numerical solution as a function of the number $N$ of elements along the fault, computed over the entire fault ([4.5,5,5]) and on 80\% of the fault ([4.6,5.4]). Convergence is observed in the interior of the fault, but not near the fault tips.

\newpage
\thispagestyle{plain} 
\mbox{}

\end{document}